\documentclass[preprint2]{aastex}

\usepackage{longtable} 
 
\usepackage{latexsym}
\usepackage{epsfig} 
\usepackage{graphicx}
\usepackage{natbib}
\usepackage{amssymb,amsbsy}
\usepackage{amsmath}
\usepackage{amsfonts}

\def\barr{\begin{array}}
\def\earr{\end{array}}
\def\berr{\begin{eqnarray}}
\def\err{\end{eqnarray}}
\def\berrno{\begin{eqnarray*}}
\def\errno{\end{eqnarray*}}
\def\be{\begin{equation}}
\def\ee{\end{equation}}

\def\fr{\frac}
\def\la{\langle}
\def\ra{\rangle}

\def\as{\prime\prime}

\renewcommand{\tt}{transformation }

\newcommand{\coos}{coordinates\,\,}

\newcommand{\pol}[1]{\stackrel{\rm LCP}{\mathrm{RCP}}}

\renewcommand{\a}{\alpha}

\newcommand{\D}{\Delta}

\newcommand{\s}{\sigma}

\shorttitle{Variables in NGC 5024}
\shortauthors{M. Safonova \& C.S.Stalin }

\begin{document}

\title{Variables in Globular Cluster NGC 5024} 

\author{M.~Safonova and C. S. Stalin}
\affil{Indian Institute of Astrophysics, Koramangala, Bangalore 560 034
    }

\begin{abstract}

We present the results of a commissioning campaign to
observe Galactic globular clusters for the search of microlensing
events. The central $10^{\prime} \times 10^{\prime}$ region of the 
globular cluster NGC 5024 was monitored using the 2-m Himalayan Chandra
Telescope in $R$-band for a period of about 8 hours on 24 March 2010. 
Light curves were obtained for nearly $10,000$ stars, 
using a modified Difference Image Analysis (DIA) technique. 
We identified all known variables within our field of view and revised 
periods and status of some previously reported short-period variables. We 
report about eighty new variable sources and present their 
equatorial coordinates, periods, light curves and possible types. Out of these, 
16 are SX Phe stars, 10 are W UMa-type stars, 14 are probable RR Lyrae 
stars and 2 are detached eclipsing binaries. Nine of the newly discovered 
SX Phe stars and two eclipsing binaries belong to the Blue Straggler 
Star (BSS) population.

\end{abstract}


\section{Introduction}

Since the beginning of the modern-day astrophysics, globular clusters (GCs) have 
been a working laboratory for observers, providing invaluable information and 
serving as models for understanding stellar dynamics. With the recent discovery 
that most of the galaxies host massive/supermassive black holes in their centres, 
the question was raised of the dynamical detection of such black holes in low-mass, 
non-active stellar systems, of which GCs are potential candidates. The well-established correlations between the properties of the supermassive 
black holes and their host galaxies do suggest that, in extrapolation, 
globular clusters follow the same relations \citep{SafonovaShastri}. 
Most of the attempts in search of the central intermediate-mass black holes 
(IMBHs) in globular clusters,however, are not 
direct and present enormous observational difficulties due to the crowding 
of stars in the GC cores. Recently, \citet{SafonovaStalin} proposed a method of 
detection of the central IMBH in GCs by microlensing (ML) of the cluster stars.
In 2010, we have initiated the observational programme to search for ML 
signatures using the 2.0-m Himalyan Chandra Telescope (HCT) at the Indian 
Astronomical Observatory (IAO), Hanle, and the 2.3-m Vainu Bappu Telescope 
(VBT) at the Vainu Bappu Observatory, Kavalur, both operated by the Indian 
Institute of Astrophysics (IIA), Bangalore. The programme consists of 
obtaining one set of observations each in $V$ and $I$ bands of a selected 
set of clusters every 15$-$20 days for the period of 5$-$7 years
\citep{SafonovaStalin}. The globular cluster NGC 5024 (M53) was observed 
as part of the commissioning observations for this programme. 

It is well known by now that any microlensing search yields a dataset suitable 
for detecting variable stars that are unrelated to ML events 
\citep[for ex.,][]{Cook2006}. Moreover, it was also discovered that the regular 
ML observations are more efficient at finding faint variables, 
being insensitive to bright ones because of saturation. In this paper, 
we report the results of our commissioning observations. This 
data was used to establish our operational procedure and to tune and test the 
data reduction pipeline. The purpose was not to specifically discover the new 
variables with these observations but rather use the data to build the 
analytical tools for use with our full time-series database. The choice to 
observe the cluster M53 was made based on convenience, as an accessible target 
at the time of observation, and on the possibility for scientific potential 
from studying the variable stars in the cluster. The details of the data set 
analyzed here are not identical to our main microlensing time-series dataset, 
but we can use these observations to test our ability to obtain high-quality 
photometry defined as light curves with low noise, both random and 
systematic, and to be able to discern astrophysical signals with the timescales 
and depth of typical GC stellar variability. 

M53 ($\a_{2000}=13^{\rm h}12^{\rm m}55^{\rm s}.3$, 
$\delta_{2000}=+18^{\circ}10^{\prime}9^{\prime\prime}$) is a moderately 
compact metal-poor with [Fe/H]=-2.04 \citep{zinn} outer halo globular cluster 
that is rich in RR Lyrae variables. Though M53 was studied extensively in the 
last decade being the second most abundant cluster in variable stars after M15,
only variables of a pulsating type have been discovered. In the latest 2010 
updated version of the catalogue of variables in M53 (Clement et al. 2001) 
there are 62 reported RR Lyrae (RRl) stars, 8 suspected long-period 
SR stars and 15 SX Phe stars. Out of almost 200 blue straggler stars (BSS) 
discovered so far in M53, \citet{Beccari08} estimated some $14\%$ to be in  
binary systems, however no eclipsing binaries were previously
found in this cluster.

Despite the high variable content of M53 and a favourable position in the galaxy 
where both field contamination and interstellar reddening are very low, 
$E(B-V)=0.02$ \citep{Schlegel}, the only extensive time-series photometry 
study has been done recently by \citet{DK09} (hereafter referred to as DK). 
Previous studies were using the point 
spread function (PSF) fitting directly to the images, which does not allow 
precise measurements in crowded fields, limiting the possibility for detecting 
small-amplitude variables. In contrast, the 
Differential Image Analysis (DIA) is a powerful technique which allows accurate 
PSF photometry even in very crowded fields. In present work, we 
apply a new pipeline based on an improved version of the differential 
imaging analysis, developed by D. Bramich (2008), to a set of 
$R$-band images in order to search for variables down to $R=21$ mag. 
We have recovered all previously known variable stars in our field of 
view and revisited all known short-period SX Phe-type stars in an 
attempt to refine their periods and coordinates. We report new candidate 
variables, determine the periods of new short-period variable stars,  
report candidate 
eclipsing binaries and flux variability amongst some of the Stetson's 
photometrically standard stars. Some of the new variables were matched to the
BSS stars discovered earlier by Beccari et al. (2008). The emphasis 
of this work is on the reporting the new short-period variable sources. 
The final characterization of the newly discovered variables, especially 
the detailed photometry of matched BSS based on their position on 
the color magnitude diagram (CMD) to estimate mass 
and temperature from isochrone fitting, will be presented in a future paper.
This paper is composed as follows. Observations and data reduction are described 
in Section~\ref{sec:2}. In Section~\ref{sec:known-var} we discuss previously 
known variables in M53, in Section~\ref{sec:new_variables} we describe the search
for new variables, explain the methods we used to identify new variable 
stars (RR Lyrae, SX Phe, eclipsing binaries), list the properties of all newly 
detected variables and display their light curves, and in Section~\ref{sec:conclusion} 
we give the summary of our results.

\section{Observations and Data Reduction}
\label{sec:2}

\subsection{Observations}
\label{sec:observations}

Photometric observations were obtained on March 24, 2010, using
the Himalayan Faint Object Spectrograph and Camera (HFOSC) mounted on the 2.0-m HCT 
of the IAO, located at 4500 m above sea level. A total of 101 image frames, 
each of 100 secs exposure, were collected in $R$-band during continuous 8 hours of 
observations. HFOSC is equipped with a Thompson CCD of $2048\times 2048$ pixels with 
a pixel scale of $0^{\as}.296$/pix and a field of view of $\sim 10'\times 10'$. 
The readout noise, gain and readout time of the CCD are 4.8 $\bar{e}$, 
1.22 $\bar{e}/{\rm ADU}$, and $ 90$ sec, respectively.

\begin{figure*}[ht!]
\begin{center}
\includegraphics[scale=1.]{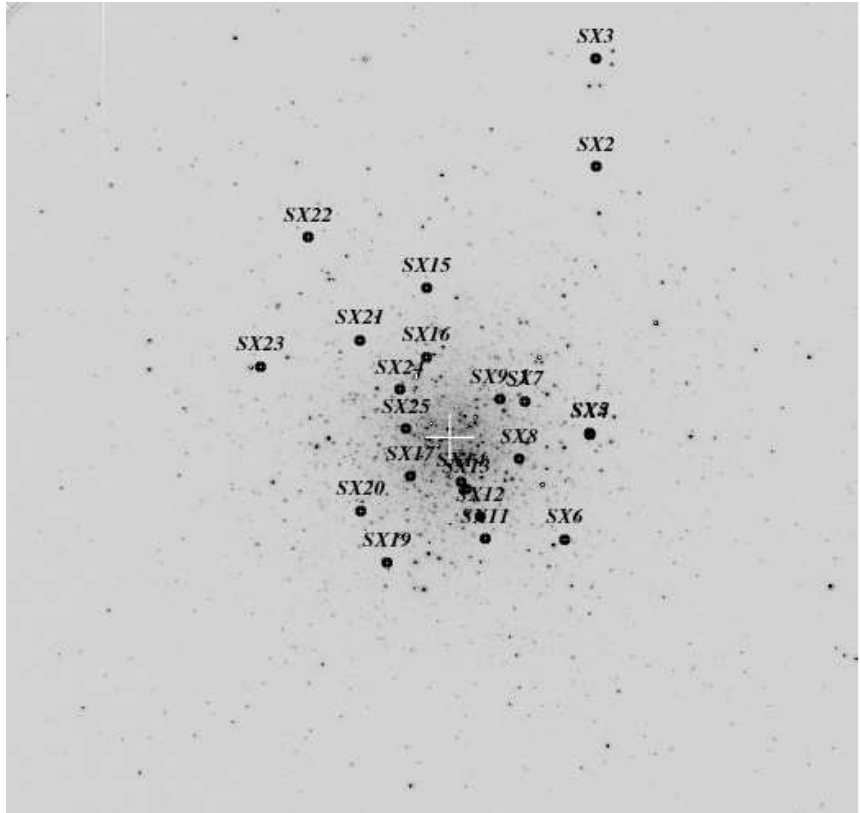}
\caption{$10\times 10$ arcmin$^2$ grey-scale map of a $R$ reference image of the globular cluster M53. 16 confirmed SX Phe stars are labelled by their respective designations given in Table~\ref{table:SXPhe}. The image was scaled 
so that to only mark the positions of the stars. The cross marks the centre 
of the cluster. The cluster size is $\sim 13^{\prime}$, thus we can be 
confident that most detected variables belong to the cluster. North is up, 
east to the left.\label{fig:SX}}
\end{center}
\end{figure*}

\begin{figure*}[ht!]
\begin{center}
\vbox{
\includegraphics[scale=0.9]{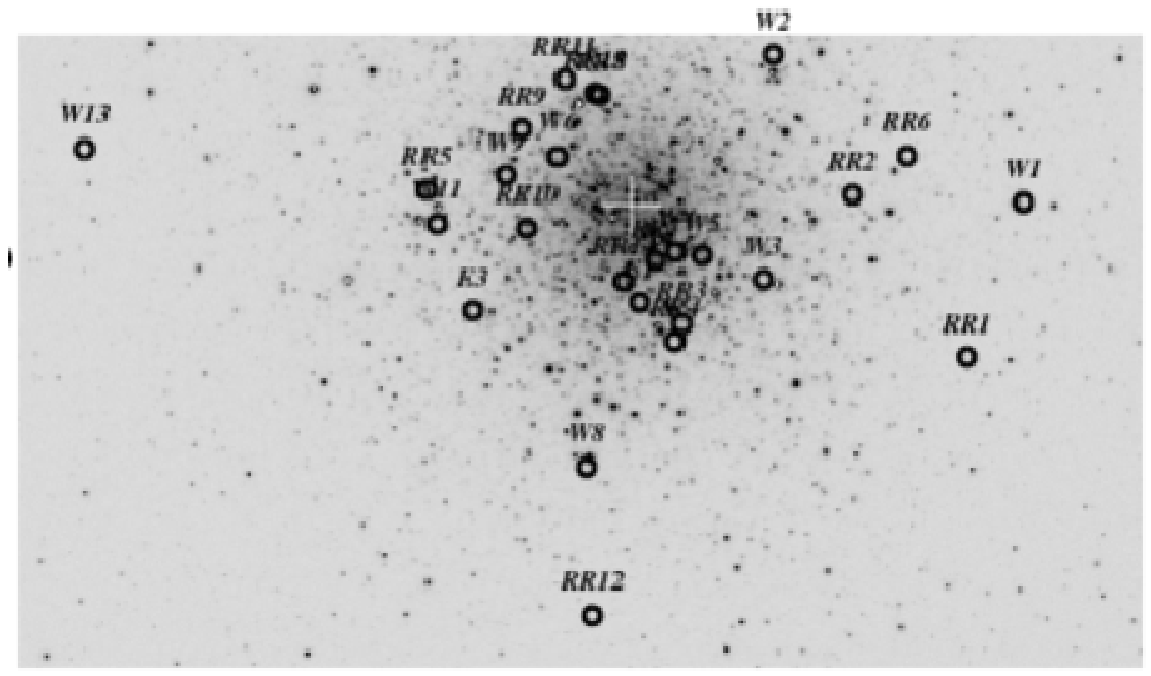}
\includegraphics[scale=0.9]{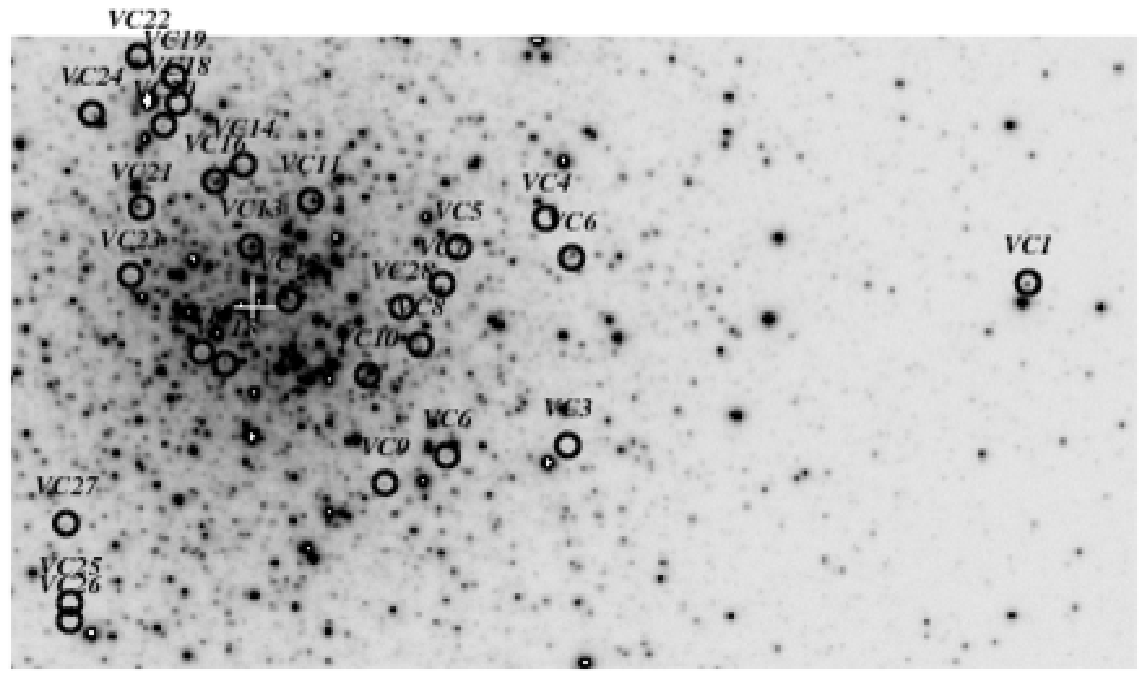}}
\caption{{\it Top}: Grey-scale map of a $R$ reference image, covering an area of about $4.0 \times  2.2$ arcmin$^2$. Variable 
candidates are labelled by their respective designations given in 
Tables~\ref{table:RRl} and \ref{tab:WUMa}. {\it Bottom}: Grey-scale map of a $R$ 
reference image, covering an area of about 
$5\times 3.5$ arcmin$^2$. 28 uncertain-type variable candidates are 
labelled by their respective designations given in 
Table~\ref{table:uncertain}. The image was cropped and 
scaled so that to only mark the positions of the 
stars. The cluster size is $\sim 13^{\prime}$, thus we can be confident 
that most variables belong to the cluster. The cross marks the centre 
of the cluster. North is up, east to the left.\label{fig:uncertainRF}}
\end{center}
\end{figure*}

A grey-scale map of a $R$ CCD image (reference frame) is shown in Figure~\ref{fig:SX}. It shows an area of about $5.5^{\prime}\times 5.5^{\prime}$ 
of the total observed field. 16 confirmed SX Phe stars are represented by circles labelled with their respective designations given in Table~\ref{table:SXPhe}.
Figure~\ref{fig:uncertainRF} {\it Top} shows the 
positions of RR Lyrae, W UMa and eclipse candidates, and {\it Bottom} of unclassified type variables, respectively. 

\subsection{Data Reduction}
\label{sec:dandia}

To extract high precision photometry from the M53 
image frames, we employed the DIA technique implemented through
a pre-release version of the pipeline DanDIA. The idea of DIA is to  
obtain information about the brightness behaviour of a source
by analyzing the difference between the image in each of the frames from the 
time-series and the image in a fixed reference frame (RF). This technique allows 
the extraction of high S/N signals even in the highly crowded central regions 
of globular clusters \citep{Alard98,Alard2000,bramich05,bramich08}. We used a 
pre-release version of DanDIA\footnote{DanDIA is build from the DanIDL library 
of IDL routines (http://www.danidl.co.uk)} pipeline \citep{bramich08}. 
This procedure as well as working of the pipeline is well described in a 
series of papers by \citet{Ferro2008}. 

Briefly, the raw image data is passed through a series of modules, starting 
with bias subtraction, flat-field corrections and cosmic rays removal. 
The gain and readout noise at the time of observations were calculated 
automatically at the first stage. An RF was chosen out of 
the best-seeing pre-processed images in which the FWHM of the PSF was 
measured to be $\sim 3.77$ pixels. A series of difference images was 
created by subtracting the RF from each registered image. Photometry on the difference images via optimal PSF scaling \citep{bramich05} have yielded 
the light curves of differential fluxes for approximately $10,000$ stars.  

\subsection{Post-processing}

\begin{figure}[ht!]
\begin{center}
\vbox{
\includegraphics[scale=0.4]{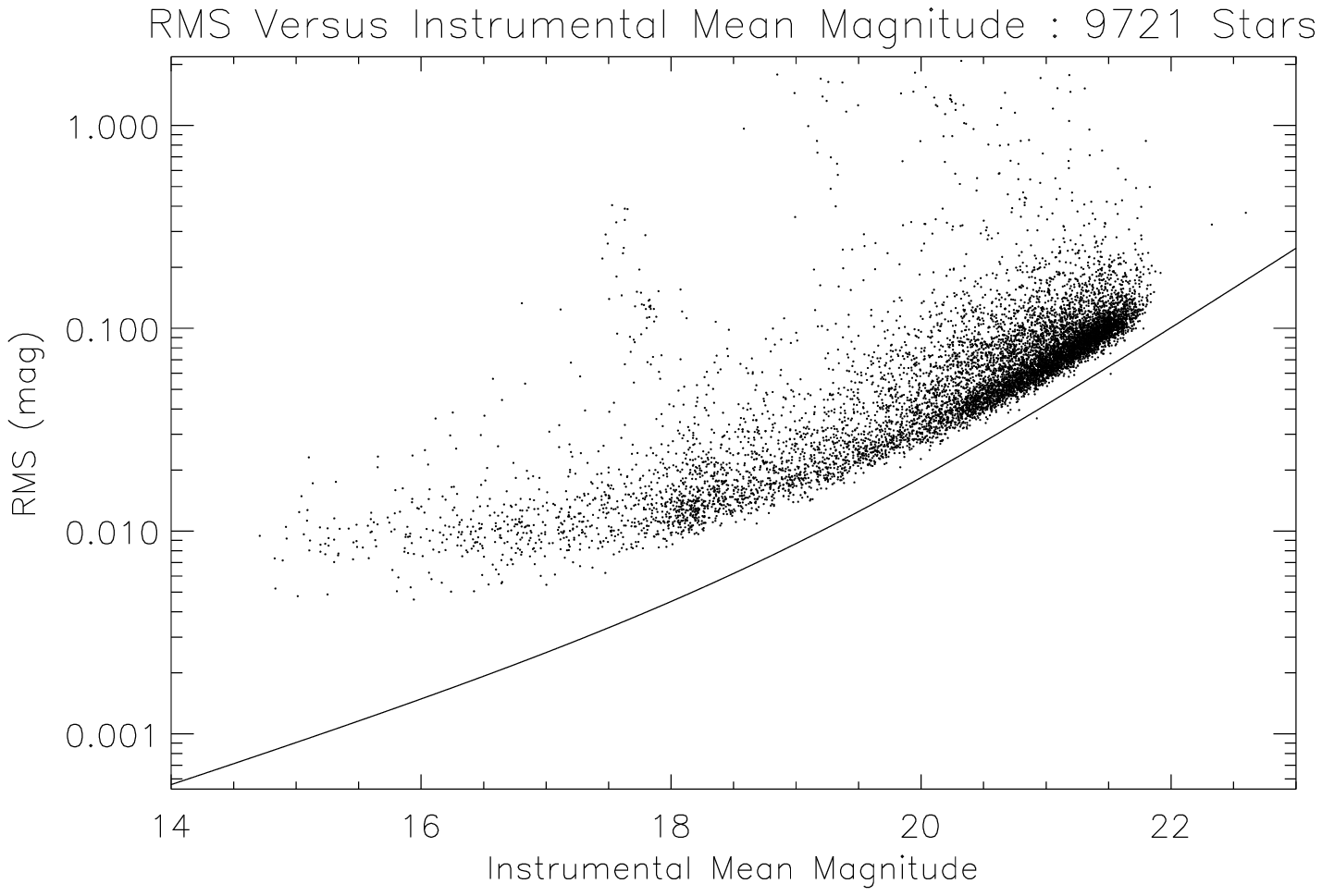}
\includegraphics[scale=0.4]{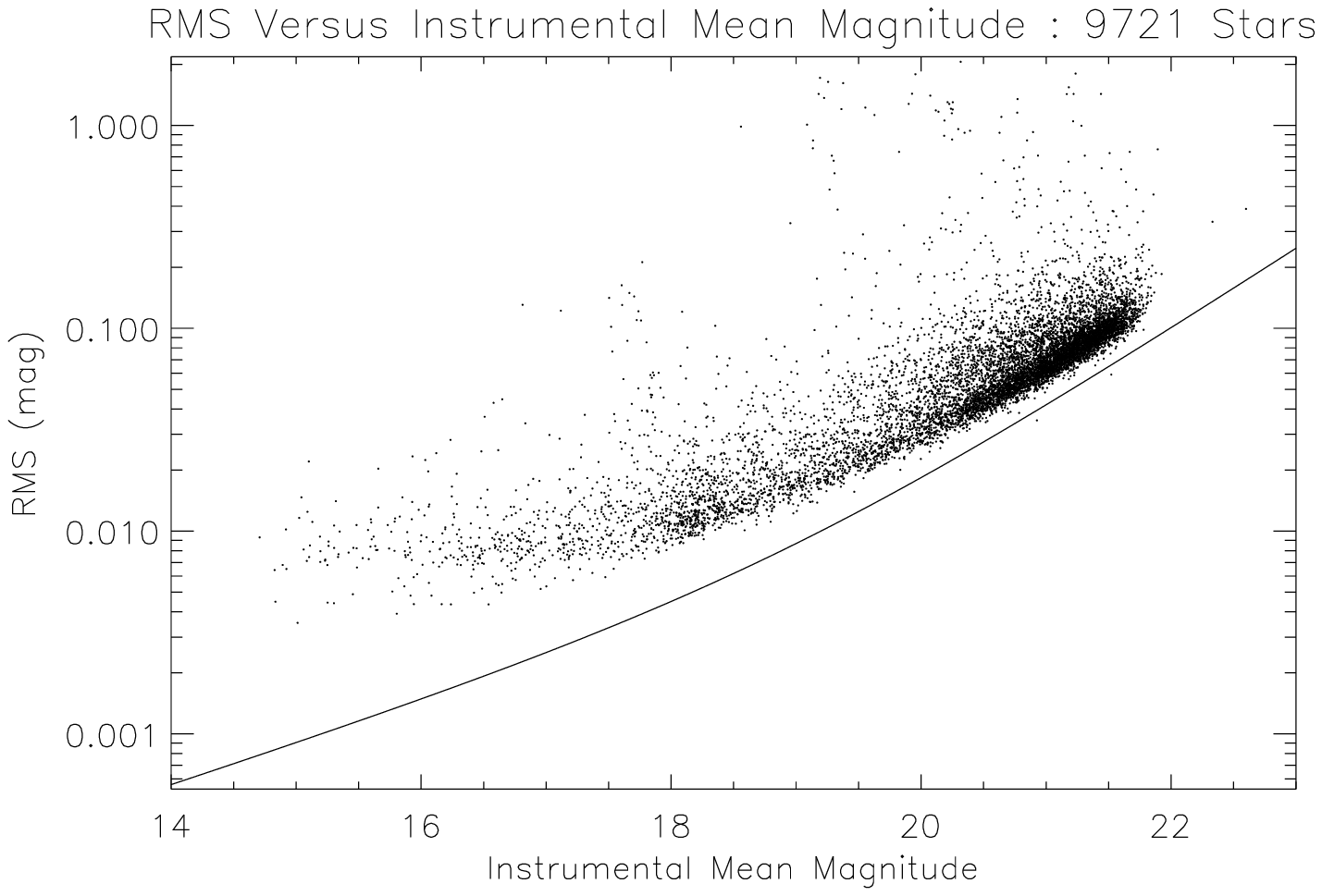}
}
\caption{{\it Top}: Plot of rms vs mean instrumental $r$ magnitude 
for each of the 9721 `raw' light curves. {\it Bottom}: Plot of rms vs mean
instrumental $r$ magnitude for each of the 9721 `post-processed' light curves. 
It is seen that there are fewer stars with large rms in the bottom plot compared 
to the top plot, and more stars with rms $< 1\%$ in the bottom plot than in 
the top one.\label{figure:rms}}
\end{center}
\end{figure}

To filter out the systematic trends caused by effects such as changes in 
seeing which may cause the spurious detection of variable 
stars, 
we pass the light curves to a final post-processing module of the pipeline, where 
the Tamuz post-calibration algorithm \citep{Tamuz} was applied. We de-trend all
images by running Tamuz algorithm for a maximum of two times, as we found that 
more than this starts to noticeably degrade signals and may remove genuine variability. 
Fig.~\ref{figure:rms} shows the rms as a function of the mean instrumental 
magnitude for 'raw' light curves and light curves after two cycles 
of post-processing. The rms frame-to-frame scatter of the instrumental magnitude 
is a good indicator of the accuracy of the photometry. In addition, stars with 
a large dispersion for a given magnitude are, in principle, good variable candidates. 
However, it is possible that a light curve has a large rms due to bad measurements 
in some images, in which case the variability is spurious. 

\subsection{Problems with Data Reduction}
\label{sec:caveats}

During the data reduction trage, we have discovered two features of 
the differential image construction with DanDIA 
that had effects on the photometry of some stars. Firstly, as noted by 
\citet{bramich2010}, due to the saturation effects on the RF, DanDIA 
discounts the area in the difference image around the saturated 
star with its immediate neighborhood. As a result, in the neighbourhood 
of saturated stars it may be impossible to extract any photometric 
measurements. We have reduced the exposure time as 
much as it is possible without degrading the S/N ratio to minimize the 
number of saturated stars. However, our images still contains several saturated 
stars, which couls have affected the photometry of nearby candidates. Consequently,
we discarded any candidate variables which were closer than 10 pixels to such stars. 

\begin{figure}[ht!]
\includegraphics[width=9cm,height=10cm]{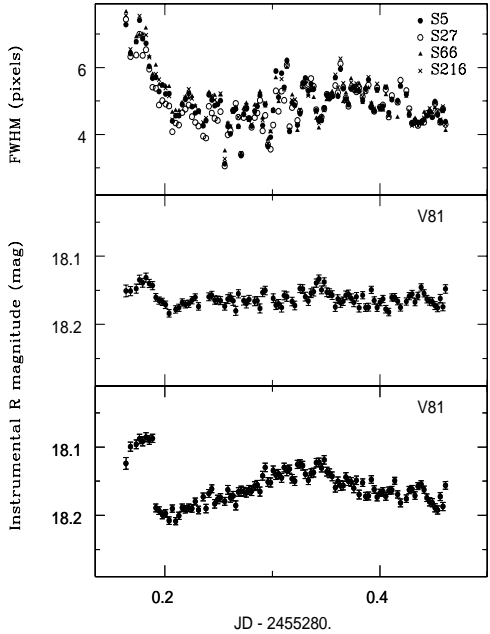}
\vskip -0.2in
\caption{{\it Top panel}: Changes of the FWHM of four Stetson's (2001) 
standard stars, {\em S5, S27, S66, S216}, with $R$ magnitudes of  
13.9, 14.5, 17.3 and 16.3 mag, respectively. {\it Middle and low
panels}: the light curves of a star {\em V81}, obtained on
running DanDIA subtraction module with $6\times 6$ and $3\times 3$ subdivision
grid, respectively. 
\label{fig:psf}}
\end{figure}

Second, at the stage of variable selection, we found that a large 
number of stars had nearly identical light curve variation which was 
correlated in time. In many cases this false variability was 
correlated with the intra-night changes in PSF. In the top panel of Figure~\ref{fig:psf} we display the changes of the FWHM (in pixels) 
of the four photometric standard stars \citep{Stetson2000} located 
at different positions on the CCD. 

We traced the origin of this effect to the combination of intra-night changes 
of the stellar PSF and the way DIA works. It is possible that this variability is  
induced at the subtraction stage of the DanDIA pipeline. When DanDIA convolves 
the RF with other frames, it uses different parameters for each subregion 
(Sec.~\ref{sec:dandia}) of the grid. It is possible that 
at the edges of these subregions DanDIA finds it difficult to fit the 
convolution parameters. For stars in those areas, the intra-night PSF changes  
mean that any consistently inadequate convolution may show up as 
photometric variability. We had re-run the subtraction stage of the pipeline
with the coarser grid of $3\times 3$ subregions and found induced changes in the
light curves of some stars. For example, star {\it V81} reported by DK, was found
by us to be non-variable on the initial run of the pipeline (Fig.~\ref{fig:psf}, {\it Middle panel}, and Sec.~\ref{sec:V81,V82,V83}). On the re-run, it started exhibiting 
a spurious variability of the type mentioned above, which is clearly seen in 
Fig.~\ref{fig:psf}, {\it Bottom panel}. It is possible that during the re-run,
this star turned out to be on the edge of the subregion while previously 
it was away from it. It is difficult to determine in advance false 
variability due to this effect, thus we have rejected such light curves through
a visual examination of variable candidates.

\subsection{Astrometry}
\label{sec:astrometry}

The astrometric transformation between pixels and celestial \coos for 
the RF was done using 46 photometric standard stars in the field of M53,
taken from P.~Stetson's 
online catalogue \citep{Stetson2000} at Canadian Astronomical Data 
Center (CADC)\footnote{The catalogue is available at  http://www3.cadc-ccda.hia-iha.nrc-cnrc.gc.ca/community/STETSON/standards/}, 
which were uniformly distributed around the cluster centre and 
located sufficiently outside the cluster core. Since our observations were 
performed at a current epoch of J2010.235873, we first precessed the coordinates 
of these standards from J2000.0 epoch to the epoch of our observation 
(using IRAF utility {\em precess}), calculated the transformation solution 
from pixel to equatorial coordinates $(\a,\delta)$ and then precessed the solution back 
to epoch J2000.0. The standard deviations in the residuals of the coordinate 
transformation were $0^{\as}.052$ and $0^{\as}.053$ in right ascension and 
declination, respectively. We have found that in 10 years (from J2000.0 to J2010.225873), 
the cluster has moved by $470^{\as}.2$ ($450^{\as}.57$ in $\a$ 
and $-194^{\as}.59$ in $\delta$). This is more or less consistent with the precession 
value of 1 degree per 72 years ($50^{\as}$/yr). The cluster's proper motion 
was estimated earlier to be very small of $\sim 0.5\pm 1$ mas/yr (Dinescu et al. 1999). 
 
\subsection{Photometric Calibration}
\label{sec:photometry}

The absolute photometric calibration was obtained by using local 
secondary standards in M53. The instrumental $r$ magnitudes were converted 
to the $R$ standard 
system by using 26 selected photometric standard stars in the field of M53 
from the Stetson's online catalogue. The $R$ magnitudes for these 
stars were obtained from the online 
USNO-A2.0 Catalogue Server (http://archive.eso.org/skycat/servers/usnoa). 

\begin{figure}[ht!]
\begin{center}
\includegraphics[width=6cm,height=6cm]{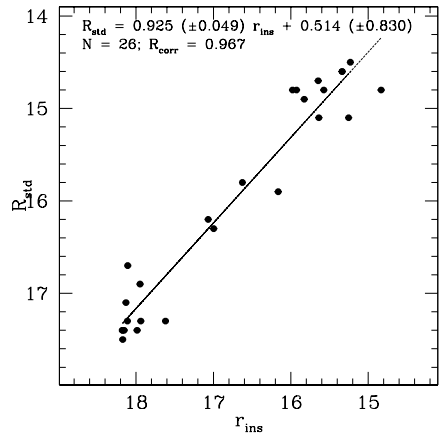}
\caption{The transformation relationship between the 
instrumental $r_{inst.}$ and the standard $R$ magnitudes obtained
using 26 secondary photometric standard stars in M53. \label{fig:mag_trans}}
\end{center}
\end{figure}

The photometric transformation equation has the fitted form 
\be
R_{\rm std}=0.925(\pm 0.049)r_{\rm inst} + 0.514(\pm 0.830)\,,
\label{eq:transformation}
\ee
where $R_{\rm std}$ is the standard magnitude and $r_{\rm inst}$ is 
the instrumental magnitude. The linear correlation coefficient is 
$R=0.967$. The linear least square fit shown in Fig.~\ref{fig:mag_trans} 
was used to obtain the standard $R$ magnitudes of new variables found 
in this study.  

\section{Variable stars in M53}

\label{sec:known-var}

According to the updated 2010 online version of catalogue by C.~Clement et al. 
(2001)\footnote{A full updatable catalogue is accessible at 
http://www.astro.utoronto.ca/people.html} there are 90 variables in the field of M53.
We have recovered the light curves of 64 of them. Out of 26 known variables which 
we could not recover, 12 were outside our field of view, 13 were saturated on our 
RF and one was at the edge of the CCD. In Table~\ref{table:var-coos}, we list the 
obtained equatorial coordinates (J2000.0) of the known variables recovered in our 
study, their offsets to the coordinates given by Clement et al. (2001) and their
2MASS identifications, if any. The offsets were calculated using
\be 
r^{\as}=\arccos\big[\sin{\delta_1}\,\sin{\delta_2}\times \cos{(\a_1-\a_2)}+
\cos{\delta_1}\times\cos{\delta_2}\big]\,,
\label{eq:offset}
\ee
where $(\a_1,\delta_1)$ and $(\a_2,\delta_2)$ are right ascensions and declinations of the 
stars for which the offset is to be calculated. We also provide the equatorial 
coordinates for variables from {\em V61} to {\em V70} which are not given in the 
Clement's catalogue. Several cases in Table~\ref{table:var-coos} deserve 
special notes and are described in the text. 

\begin{deluxetable}{cccccl}
\tablecaption{\small Revised equatorial coordinates of identified known 
variables in M53 and their offsets with those given in the catalogue by 
Clement et al (2001).\label{table:var-coos}} 
\tablewidth{0pt}
\tablehead{
\colhead{ID}     & \colhead{$\a$(2000)} & \colhead{$\delta$(2000)}       &
\colhead{Offset} & \colhead{2MASS ID \#}   & \colhead{Notes/Remarks}\\
\colhead{ }      & \colhead{(h:m:s)}    & \colhead{($^{\circ}:':''$)} &
\colhead{$^{''}$} & \colhead{ }          & \colhead{ }
}
\startdata
V1  &13:12:56.34 & 18:07:13.8 & 0.823 &  J13125634+1807139  &  \\
V2  &13:12:50.27 & 18:07:00.8 & 0.511 &  J13125028+1807007  &  \\
V3  &13:12:51.37 & 18:07:45.4 & 0.688 &  J13125132+1807461  &  \\ 
V4  &13:12:43.88 & 18:07:26.2 & 0.559 &  J13124390+1807259  &  \\
V5  &13:12:39.09 & 18:05:42.6 & 0.349 &  J13123907+1805424  &  \\
V6  &13:13:03.90 & 18:10:20.1 & 0.541 &  J13130390+1810202  &  \\
V7  &13:13:00.85 & 18:11:30.0 & 0.609 &  J13130086+1811302  &  \\
V8  &13:13:00.42 & 18:11:05.1 & 0.615 &  J13130043+1811049  &  \\
V9  &13:13:00.07 & 18:09:25.1 & 0.639 &  J13130013+1809256  &  \\
V10 &13:12:45.72 & 18:10:55.5 & 0.467 &  J13124572+1810554  &  \\
V11 &13:12:45.37 & 18:09:01.9 & 0.583 &  J13124540+1809020  &  \\
V15 &13:13:12.374& 18:13:55.6 & 0.500 &           -         &   \\
V16 &13:12:46.19 & 18:06:39.1 & 0.541 &  J13124620+1806387  &  \\
V17 &13:12:40.34 & 18:11:54.1 & 0.418 &  J13124030+1811540  &   \\
V18 &13:12:48.67 & 18:10:13.1 & 0.590 &  J13124868+1810128  &  \\
V19 &13:13:07.00 & 18:09:26.3 & 0.559 &  J13130702+1809262  &  \\
V22 &13:12:51.97 & 18:05:16.7 & 0.327 &  J13125197+1805165  &  \\
V23 &13:13:02.33 & 18:08:36.1 & 0.443 &  J13130235+1808359  &   \\
V24 &13:12:47.28 & 18:09:32.3 & 0.566 &     -               &   \\
V25 &13:13:04.40 & 18:10:37.3 & 0.664 &  J13130441+1810372  &   \\
V27 &13:12:41.42 & 18:07:23.7 & 0.541 &  J13124143+1807235  &   \\
V29 &13:13:04.26 & 18:08:47.0 & 0.443 &  J13130426+1808468  &   \\
V31 &13:12:59.56 & 18:10:04.9 & 0.535 &  J13125960+1810046  &   \\
V32 &13:12:47.72 & 18:08:35.8 & 0.488 &  J13124774+1808358  &   \\
V33 &13:12:43.89 & 18:10:12.1 & 1.416 &       -             & see individual notes  \\
V34 &13:12:45.70 & 18:06:26.0 & 0.566 &  J13124569+1806258  &   \\
V35 &13:13:02.41 & 18:12:37.9 & 0.464 &  J13130235+1812373  &   \\
V36 &13:13:03.28 & 18:15:10.5 & 0.535 &  J13130329+1815109  &   \\
V37 &13:12:52.28 & 18:11:05.3 & 0.492 &  J13125227+1811050  &   \\
V38 &13:12:57.14 & 18:07:40.5 & 0.727 &  J13125713+1807404  &   \\
V39 &13:12:38.74 & 18:13:31.3 & 0.516 &  J13123874+1813312  &   \\
V40 &13:12:55.85 & 18:11:54.8 & 0.535 &  J13125583+1811545  &   \\
V41 &13:12:56.75 & 18:11:08.5 & 0.630 &  J13125679+1811093  &   \\
V42 &13:12:50.55 & 18:10:20.0 & 0.395 &  J13125058+1810195  &   \\
V43 &13:12:53.08 & 18:10:55.4 & 0.511 &  J13125306+1810552  &   \\
V44 &13:12:51.66 & 18:10:00.5 & 0.590 &       -             &    \\
V45 &13:12:55.15 & 18:09:27.4 & 0.590 &  J13125517+1809274  &   \\
V46 &13:12:54.53 & 18:10:37.1 & 0.418 &          -          &    \\
V47 &13:12:50.41 & 18:12:24.7 & 0.492 &  J13125043+1812246  &   \\
V51 &13:12:57.58 & 18:10:49.6 & 0.549 &      -              &     \\
V52 &13:12:55.91 & 18:10:37.0 & 0.327 &      -              & see individual notes \\
V53 &13:12:55.78 & 18:10:36.0 & 0.615 &  J13125580+1810360  & see individual notes \\
V54 &13:12:54.29 & 18:10:31.5 & 0.590 &     -               &  \\
V55 &13:12:53.45 & 18:10:36.6 & 0.492 &      -              &    \\
V56 &13:12:53.69 & 18:09:26.0 & 0.590 &      -              &   \\
V57 &13:12:55.44 & 18:09:57.9 & 1.699 &  J13125547+1809577  &BSS, see individual notes\\
V58 &13:12:55.59 & 18:09:31.0 & 0.492 &     -               &   \\
V59 &13:12:56.67 & 18:09:20.8 & 0.441 &     -               &    \\
V60 &13:12:56.99 & 18:09:36.5 & 0.418 &  J13125695+1809357  &   \\
V61-V70&         &            &       &                     & see individual notes\\
V72 &13:12:55.942& 18:09:52.12& 2.522 &                     & see individual notes \\
V73 &13:13:03.34 & 18:09:25.1 & 0.655 &     -               &    \\
V74 &13:12:49.68 & 18:07:25.9 & 0.427 &     -               &BSS, \citet{Beccari08}\\
V75 &13:13:09.39 & 18:09:39.7 & 0.792 &     -               &BSS, first report  \\
V76 &13:13:04.97 & 18:08:35.8 & 0.658 &     -               &BSS, \citet{Beccari08}\\
V79 &13:12:46.60 & 18:11:36.7 & 0.148 &     -               &BSS, DK   \\
V80 &13:12:57.46 & 18:10:14.8 & 1.294 &     -               & see individual notes \\
V81 &13:13:02.69 & 18:06:29.7 & 0.283 &  J13130271+1806294  &   \\
V82 &13:12:56.44 & 18:13:09.9 & 0.850 &     -               &    \\
V83 &13:12:50.11 & 18:07:43.0 & 0.187 &     -               &   \\
V87 &13:13:01.92 & 18:10:13.2 & 0.475 &     -               &BSS, first report \\
V89 &13:13:08.15 & 18:07:38.4 & 0.884 &     -               &BSS, first report \\
\enddata
\tablecomments{The coordinates listed above for all variables numbered up to 
{\em V60} are from \citet{Evstigneeva}. Coordinates for {\em V73}--{\em V76}, 
{\em V87} and {\em V89} are from \citet{Jeon03}; for variables 
{\em V79}--{\em V83} from DK.} 
\end{deluxetable}

\subsection{Variable stars with periods $P> 0.1$ days}
\label{sec:LPV}

Though we have obtained light curves for most of previously detected variables
in this period range, we would not like to make conclusions regarding their
variability due to the short span of our observations. However, several stars in this
period range deserve special mention.

\subsubsection{Notes on individual variables}

\subsubsubsection{V33} Both Clement's (obtained from Evstigneeva 
et al. 1997) and our coordinates do not match with the position of this 
variable as 
marked in Kopacki (2000) ID chart. However, it is situated very close to a 
very bright secondary standard star {\em S4} that is nearly saturated on our RF, 
which may account for the shift in the light centroid in our case.

\subsubsubsection{V52-V53}

These two stars separated by $\sim 2^{\as}$ and DK reported that they could 
not resolve them. However, {\em V52} is clearly resolved on our images and 
its \coos coincide with the \coos given by Clement et al. (2001) with $0.^{\as}3$ 
offset. Though we could not determine its period, its light curve shows the 
RRlab-type variability. At the position of {\em V53}, given by Kopacki (2000), 
there are 3 stars, one of which does show some variability; however its 
independent variability is under question. 2MASS position for {\em V53} 
is between these three stars. 

\subsubsubsection{V57}
\label{subsec:V57}

\begin{figure}
\vbox{
\vspace*{-2.0cm}
\hspace*{1.0cm}\includegraphics[scale=0.5]{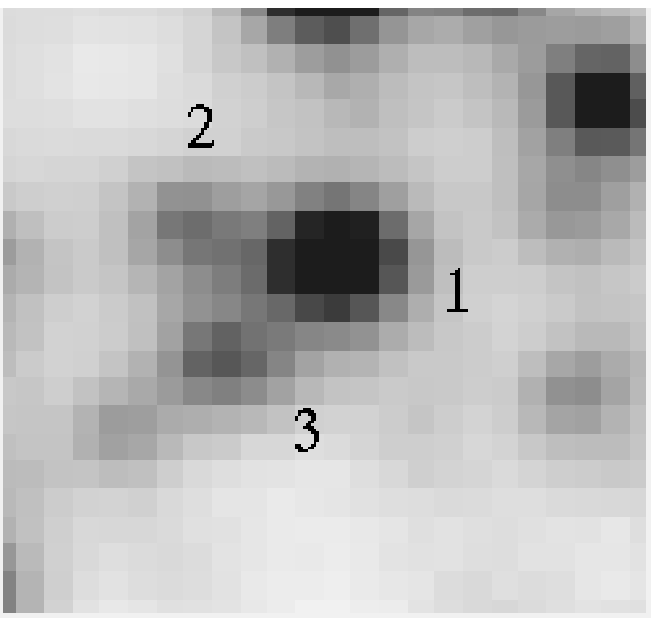}
\hspace*{-1.0cm}\includegraphics[width=9cm,height=7cm]{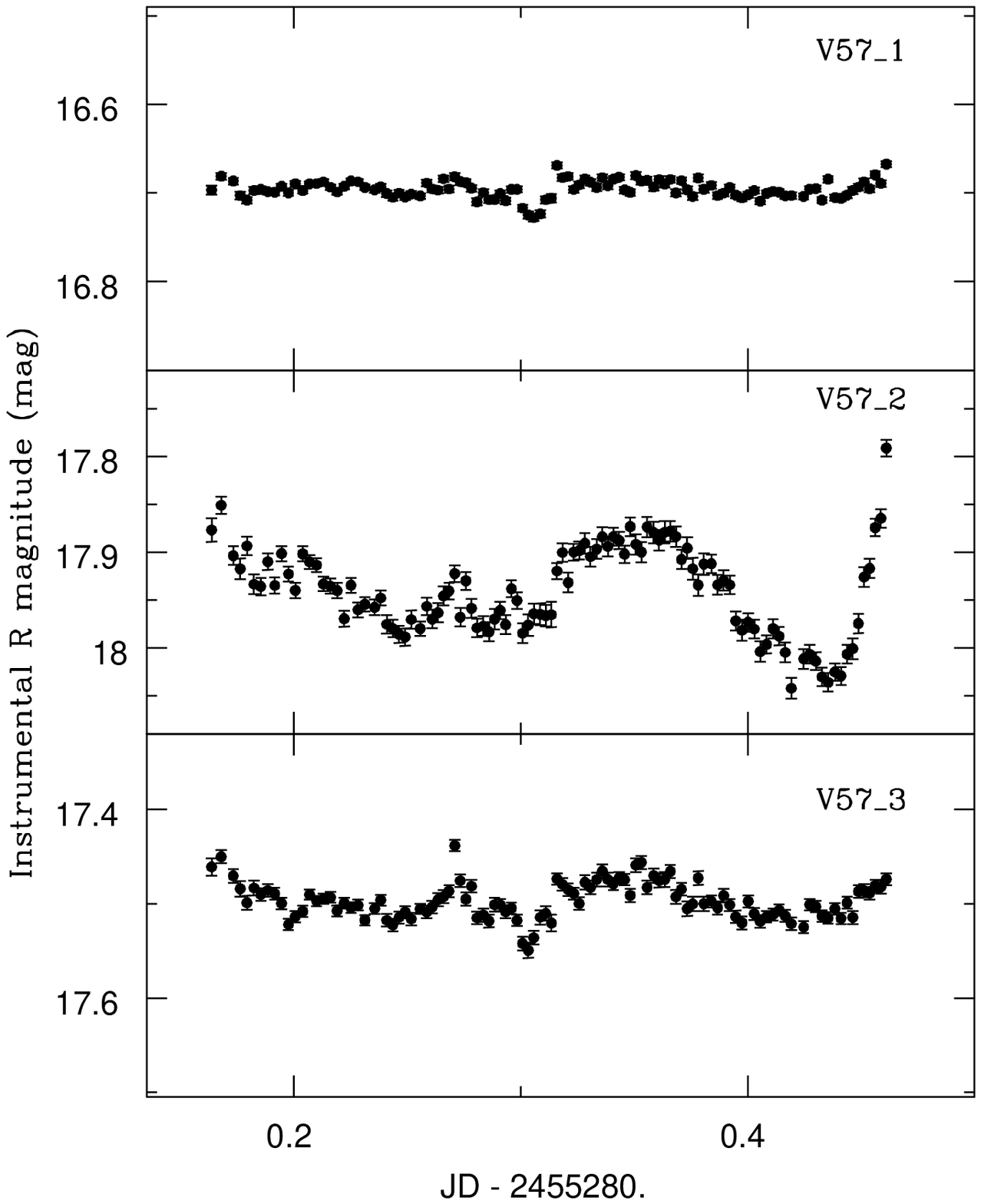}
\vskip -0.3in
\caption{$7^{\prime\prime}\times 7^{\prime\prime}$ field around 
{\em V57} with marked stars and the light curves 
of {\em V57\_1}, {\em V57\_2} and {\em V57\_3}. The names of the stars 
are indicated on each panel. \label{fig:V57}}
}
\end{figure}

The coordinates of a star, identified by Kopacki (2000) as {\em V57}, do not 
match with the coordinates given by Evstigneeva et al. (1997). Coordinates of
Evstigneeva et al. (1997) match another, fainter, star close to it. To 
investigate further, we have checked both stars, marked {\em V57\_1} and 
{\em V57\_2}, respectively, for variability. The star identified by 
Kopacki as {\em V57} (our {\em V57\_1}) does not 
show any variability, though with the listed period of $P=0.5683$ days, its light 
curve should have shown some variation over our 7.5 hrs of observations. 
Whereas the star that matches with Evstigneeva's coordinates (our {\em V57\_2}) 
shows very 
clear variability with possible two pulsation periods of different amplitudes. 
Due to the limited time span of our observations, we could not determine its 
period, but the presence of Blazhko effect can be suspected from the observed 
light curve. Thus, we conclude that the star identified and marked on the ID 
chart by Kopacki (2000) as {\em V57} is a mis-identification, and the correct 
coordinates are given in Evstigneeva et al. (1997). Interestingly, there is 
one more star within this field, marked as {\em V57\_3}, which shows 
complicated variability of $\sim 0.2$ d and a small amplitude of $\D m 
\sim 0.11^m$. In see Fig.~\ref{fig:V57} we show all three stars, marked accordingly, 
and their time-domain light curves. When we cross-correlated our 
list of known variables with the BSS catalogue by Beccari et al. (2008) 
(see Sec.~\ref{sec:BSS}), we have found a match of {\em V57} with the BSS 
\#102387 from HST/WFPC2/PC sample to within $0^{\as}.6$. Since {\em V57} 
was identified as a RR Lyrae previously \citep[for ex.,][]{Kopacki2000}, 
it is possible that it is {\em V57\_3} that is the BSS 
identified by \citet{Beccari08}. More work on a identification is in 
progress and the result will be reported in the forthcoming paper.

\subsubsubsection{V61-V70}

For these stars the updated 2010 online version of catalogue by Clement et al. 
(2001) does not provide equatorial coordinates. We have determined the
coordinates of all of these stars and matched them to the 2MASS catalogue 
(Table~\ref{table:coordinates61-70}). For {\em V61} DK did not provide the light
curve as they stated that this star is merged with the long-period variable 
{\it V49} on their images. However, it shall be noted that variable 
{\em V49} (2MASS ID\# J13125915+1814356) is a star well outside the 
cluster core at a distance of $4^{\prime}.54$ from the cluster centre, 
while {\em V61} is well inside the core at only $1^{\as}.8$ 
from the centre; thus there is some mistake in their identification.  

We shall note that for variables {\em V62}, {\em V63} and {\em V64} three sets of 
coordinates are available: from the online catalogue by Samus et al. 
(2009)\footnote{\footnotesize The full catalogue is available at http://vizier.cfa.harvard.edu/viz-bin/VizieR?-source=J/PASP/121/1378. 
However, there is a mistake in the online catalogue star listing --- the \coos of a 
variable V46 are lost, instead the \coos of V47 are assigned to it, and 
this mistake carries on till the end of the catalogue.}, derived 
from 2MASS catalogue (probably precessed from J2000.0 to J2000.343), 
DK's and ours. For {\em V62}, Samus et al. (2009) give a position between {\em V62} 
and a star to the north from it with $0.^{\as}535$ offset, while DK's and 
our \coos coincide within $0^{\as}.12$ with its position. {\em V63} has no 
2MASS match; Samus et al. (2009) and our \coos match within $0.^{\as}02$, 
but DK \coos are off by $2^{\as}.44$ and actually mark a different, 
non-varying star. For {\em V64}, Samus et al. (2009) use 2MASS 
\coos (at J2000.343 epoch), but these give a position shifted by 
$0^{\as}.374$ to the left, and DK miss the position by $1^{\as}.215$. 
The remaining variables, {\em V65} to {\em V70}, match perfectly with 
2MASS sources. {\em V65} is a Stetson's (2000) secondary standard star {\em S67} 
(see Sec.~\ref{sec:stan-var}).

\begin{deluxetable}{ccccc}
\tablecaption{\small Equatorial coordinates ($\a,\delta$) of known 
variables {\em V61-V70}.\label{table:coordinates61-70}}
\tablewidth{0pt}
\tablehead{
\colhead{ID}  & \colhead{$\a$(2000)} & \colhead{$\delta$(2000)} & 
\colhead{2MASS ID \#(J2000.0)} & \colhead{Note} \\
\colhead{ }  & \colhead{(h:m:s)} & \colhead{($^{\circ}:':''$)}   &
\colhead{ } & \colhead{ }
}
\startdata
V61 & 13:12:55.214 & 18:10:10.31 & J13125521+1810103& \\
V62 & 13:12:53.995 & 18:10:29.81 & J13125400+1810302& \\
V63 & 13:12:56.300 & 18:10:00.75 & - & \\
V64 & 13:12:52.515 & 18:10:12.54 & J13125254+1810125& \\
V65 & 13:13:04.669 & 18:10:59.44 & J13130467+1810594& $\equiv$S67\\
V66 & 13:13:01.578 & 18:10:03.90 & J13130157+1810039& \\
V67 & 13:13:01.024 & 18:10:09.50 & J13130102+1810095& \\
V68 & 13:12:56.577 & 18:08:23.13 & J13125657+1808231& \\
V69 & 13:12:55.160 & 18:10:19.40 & J13125616+1810194& \\
V70 & 13:12:55.320 & 18:09:42.00 & J13125532+1809420& \\
\enddata
\end{deluxetable}

\subsubsubsection{V72}

\begin{figure}[hb!]
\begin{center}
\vbox{
\includegraphics[scale=0.8]{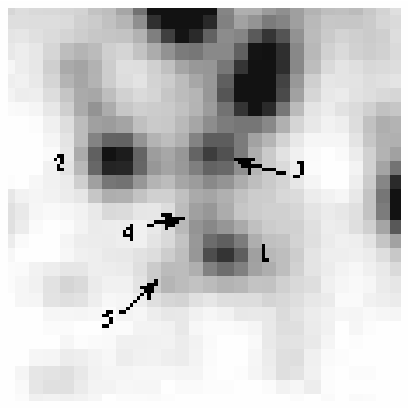}
\vskip -0.1in
\includegraphics[width=9cm,height=9cm]{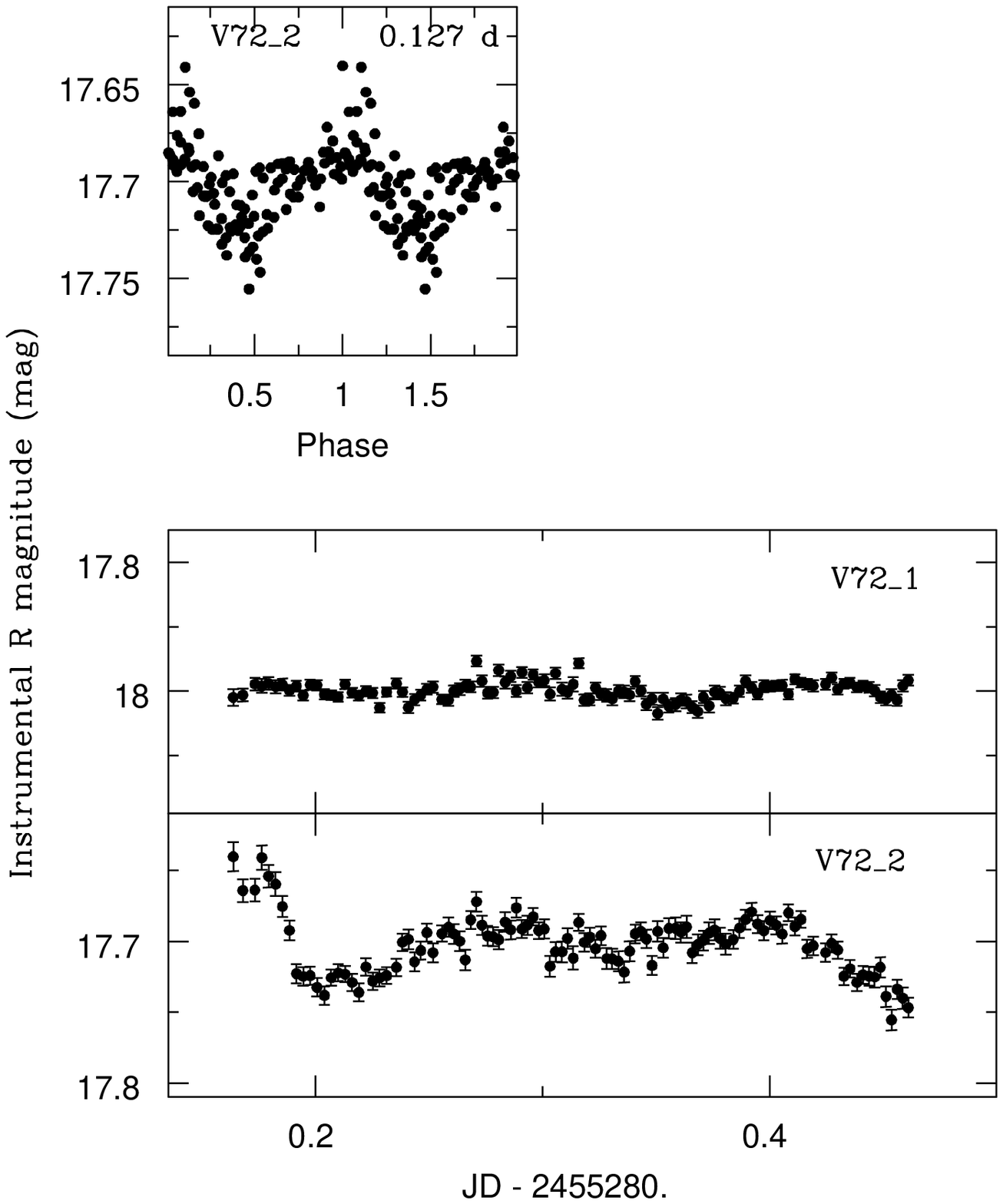}
\vskip -0.3in
\caption{$6.^{\prime\prime}5\times 6.^{\prime\prime}5$ 
field around {\em V72} position as given by DK with marked stars, 
phase plot of {\em V72\_2} and light curves of {\em V72\_1} and 
{\em V72\_2} stars.\label{figure:V72}}
}
\end{center}
\end{figure}

We did not find any variability of the star that matches the coordinates given by 
DK. Though {\em V72} is situated in the heavily crowded core, we have clearly 
identified five stars within $2^{\as}.75$ radius of the coordinates given
by DK. The star which is variable according to DK is
marked as {\em V72\_1}, the rest as {\em V72\_2}, {\em V72\_3}, 
{\em V72\_4} and {\em V72\_5}, respectively. While four of them do not show obvious  
variability, the star {\em V72\_2} is clearly varying with an estimated period of 
 0.12710 days and an amplitude of $\sim 0^m.1175$. We conclude that it 
is the star {\em V72} identified by DK, and that the period reported by them 
$P_{\rm DK}=0.2542$ days is a multiple of the true period. The coordinates of 
{\em V72\_2} and its offset from DK \coos are given in Table~\ref{table:var-coos}.
In Fig.~\ref{figure:V72} we show the $6^{\as}.5\times 6^{\as}.5$ 
field around the \coos given by DK, phase plot of {\em V72\_2} and time-domain 
light curves of these two stars.
 
\subsection{Variables with periods $P< 0.1$ days}
\label{sec:SPV}

There are 15 short-period variable (SPV) stars of SX Phe type reported so far
in M53. From our observations we were able to recover 11 of those
stars. Three stars {\em V77}, {\em V88} and {\em V90}, 
were out of our FOV and one, {\em V78}, was saturated on our RF. For these 11 
stars we have obtained the light curves and determined the periods. In several 
cases we have revised the periods given previously in the literature. 
In Table~\ref{table:SPV}, we list these stars along with the previously
reported periods, new periods and the average $R$-band
magnitudes determined in this work.

\begin{deluxetable}{lcccl}
\tablecaption{\small Revised periods for previously known 
short-period variables in M53. Column (1) is star's ID by 
Clement et al. (2001) nomenclature, column (2) -- periods 
from the literature, columns (3) and (4) -- new periods and 
standard $R$ magnitudes found in this work.\label{table:SPV}}
\tablewidth{0pt}
\tablehead{
\colhead{ID}  & \colhead{Period}  & \colhead{New period} &
\colhead{$<R>$} & \colhead{Note}\\
\colhead{ }  & \colhead{(days)}  & \colhead{(days)} &
\colhead{(mag)} & \colhead{ }
}
\startdata
V73     & 0.0701 & 0.071530   & 18.927  &       \\
V74     & 0.0454 & 0.045055   & 19.054  &   BSS   \\
V75     & 0.0442 & 0.044178   & 19.455  &   BSS      \\
V76     & 0.0415 & 0.041467   & 19.434  &   BSS      \\
V79     & 0.0463 & 0.046255   & 19.183  &   BSS      \\
V80     & 0.0674 & 0.065668   & 17.915  &   see individual notes  \\
V81     & 0.0714 &    NV      & 17.310  &   see individual notes \\
V82     & 0.0221 &    NV      & 18.864  &   see individual notes \\
V83     & 0.1247 &    NV      & 18.749  &   see individual notes \\
V87     & 0.0479 & 0.046855   & 19.356  &   BSS \\
V89     & 0.0435 & 0.43278    & 19.435  &   BSS      \\  
\enddata
\tablecomments{Periods in the second column are from DK, except
for the stars {\em V87} and {\em V89} where periods are from Jeon et al. 
(2003). NV stands for non-variable and BSS means that the star belongs
to the BSS population.}
\end{deluxetable}

\subsubsubsection{V80}
\label{subsec:V80}

\begin{figure}[h!]
\vbox{
\hspace*{2.0cm}\includegraphics[scale=0.6]{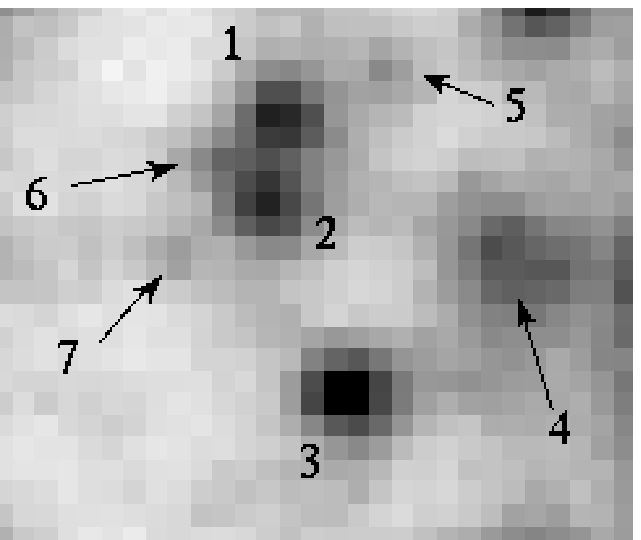}
\hspace*{-1.0cm}
\includegraphics[width=10cm,height=10cm]{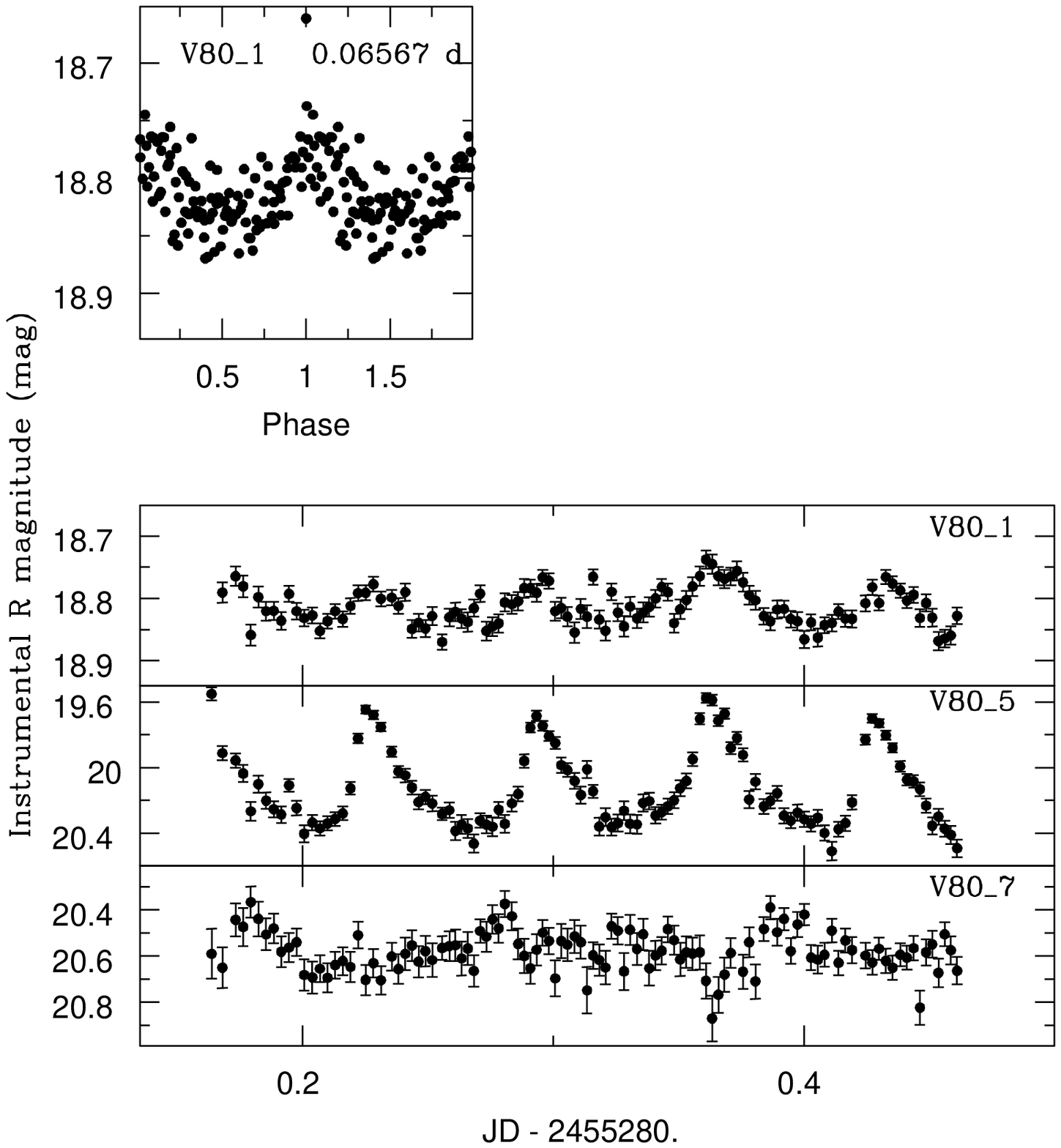}
}
\vskip -0.4in
\caption{$6.^{\prime\prime}5\times 6.^{\prime\prime}5$ 
field around {\em V80} with marked stars, phase plot of {\em V80\_1} and 
light curves of {\em V80\_1}, {\em V80\_5} and {\em V80\_7} 
stars.\label{fig:V80}}
\end{figure}

At coordinates given by DK there is no star. However, within the circle 
radius $3^{\as}.184$ of this location there are 7 stars clearly seen on our 
RF. This image region is shown in Fig.~\ref{fig:V80} with 7 stars marked on it. 
We have run the periodicity check (see details in Sec.~\ref{sec:LS}) on 
the light curves of all of these stars. Three of them, {\em V80\_1}, 
{\em V80\_5} and {\em V80\_7}, 
show short-term variability and their light curves are shown 
in Fig.~\ref{fig:V80}. Of these, {\em V80\_1} and {\em V80\_5}, show similar 
light curves and periods characteristic of SX Phe stars, but {\em V80\_5} 
has much larger variation amplitude of $\D m=0.9$ compared with $\D m=0.2$ 
of {\em V80\_1}. The light curve of a variable {\em V80\_7} is noisy with an 
amplitude of $\D m\approx 0.6$. It shall be noted that 
DK could not have determined the exact position of {\em V80} as they reported 
that this particular field was heavily crowded or blended on their RF. By 
comparing the light curves of their {\em V80} and our candidates, we conclude 
that the most probable match is candidate {\em V80\_1} (it has the same symmetric 
sinusoidal curve). The other two candidates, {\em V80\_5} and {\em V80\_7}, 
thus constitute new variables. Details on {\em V80\_5} are presented 
in Table~\ref{table:SXPhe}, where it is assigned a name {\em SX25}, 
and on {\em V80\_7} in Table~\ref{table:uncertain}. 

\subsubsubsection{V81, V82, V83}
\label{sec:V81,V82,V83}

\begin{figure}[h!]
\hspace*{-1.0cm}
\includegraphics[width=10cm,height=8cm]{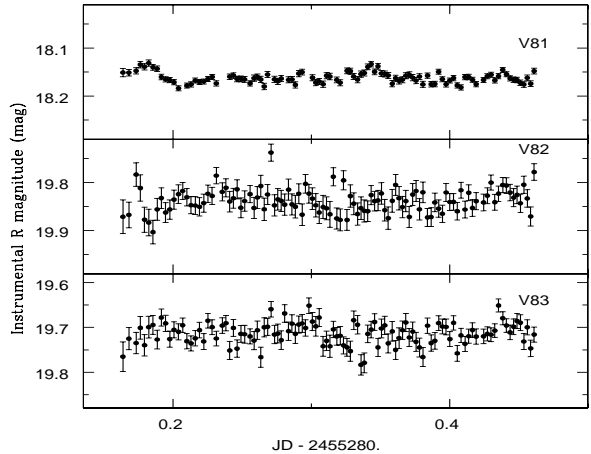}
\vskip -0.3in
\caption{Light curves of {\em V81}, {\em V82} and {\em V83} stars.\label{fig:V81}}
\end{figure}

These three stars was first reported by DK as SX Phe stars. {\em V81} 
star is clearly resolved on our RF and has a 2MASS match. 
However, we find it non-variable. DK have reported that its variability 
is of either unknown type or that it was blended with a nearby suspected 
BSS, identified by them with USNO star B1.0 1081-0245846. However, 
USNO coordinates are located between {\em V81} and two nearby faint stars 
which do not show any variability. Stars {\em V82} and {\em V83} are also 
clearly resolved on our RF and also do not show variability of the type 
reported by DK (see Fig.~\ref{fig:V81}). The results of the variability 
criteria (details are in Sec.~\ref{sec:criteria}) shown in Table~\ref{table:V81} 
also indicate that these three stars are most probably not variable.

\begin{deluxetable}{cclcc}
\tablecaption{\small Results of the variability criteria for stars 
{\em V81}, {\em V82} and {\em V83}. ${\cal A}$ is the alarm statistics, 
${\cal F}$ is the significance level of periodicity found, rms is the 
standard deviation of the mean instrumental magnitude and $\sigma_{XS}$ 
is the excess variance.\label{table:V81}}
\tablewidth{0pt}
\tablehead{
\colhead{ID}  &  \colhead{${\cal A}$}  & \colhead{${\cal F}$} & 
\colhead{rms} &  \colhead{$\sigma_{XS}$}
}
\startdata
V81 & 2.7      & $5.084 \times 10^{-3}$ & 0.01123  & 0.05233\\
V82 & -0.09842 & $7.49\times 10^{-2}$  & 0.02073  & 0.07512\\
V83 & 0.45415 & $7.044\times 10^{-1}$  & 0.01923  & 0.08208 \\
\enddata
\end{deluxetable}

\subsection{Standards as variables}
\label{sec:stan-var}

Out of 192 secondary photometric standard stars known in M53 we have selected 
43 for our astrometric calibration (Sec.~\ref{sec:astrometry}). However, 
when we tried to use these stars to devise the variable search criteria 
in this work, we found that some of the standard stars  
exhibit some sort of variability over the course of 
our ~7.5 hrs of observations, ranging from slow variations in magnitude to 
sudden aperiodic changes. This is not the first time when standard stars in 
this cluster are found to be variable. Two variables discovered by DK, 
namely {\em V84} and {\em V85}, and identified by them as LPV with periods 
of 22.4 and 19.8 days, respectively, are in fact the secondary photometric 
standard stars {\em S1} and {\em S17} of Stetson (2000). One more variable 
star {\em V65} identified by Kopacki (2000) as SR type, although 
he couldn't determine its period, is in fact the standard 
star {\em S67} from Stetson (2000). We, however, could not confirm 
their variability as all three stars were saturated on our RF.

\begin{figure}[ht!]
\includegraphics[width=8cm,height=10cm]{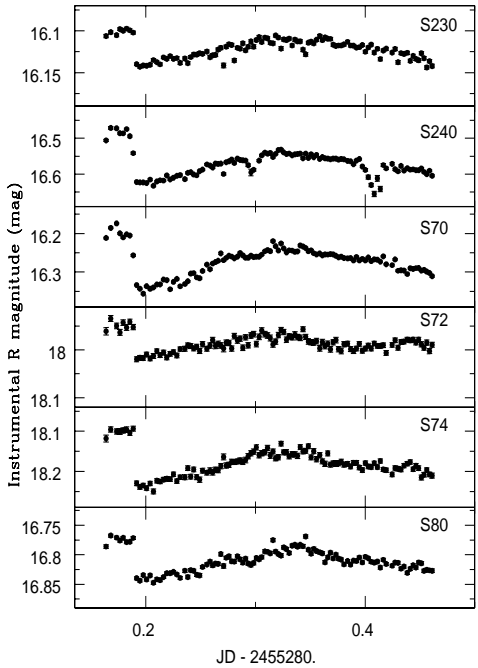}
\vskip -0.3in
\caption{Light curves of standards suspected as variable.\label{fig:spurious}}
\end{figure}

\begin{figure}[ht!]
\hskip -0.2in
\includegraphics[width=10cm,height=10cm]{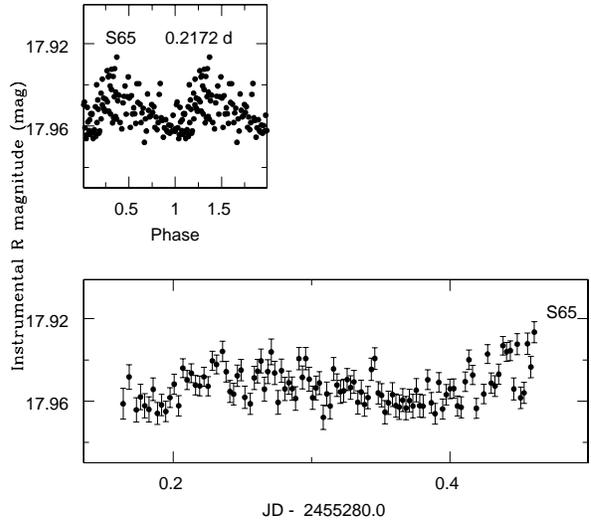}
\vskip -0.3in
\caption{Phase and time-domain curves of the photometric standard star
{\em S65}.\label{fig:S65}}
\end{figure}

We have examined the variability in our set of selected standards. The properties 
of some of these stars are given in Table~\ref{table:var-standards} and 
their light curves in Fig.~\ref{fig:spurious}. We noticed that several stars 
exhibit nearly identical variability and have concluded that this is the result 
of spurious, or induced, variability due to the combination of the intra-night 
PSF changes and DanDIA reduction procedure (see Sec.~\ref{sec:caveats}). Out 
of these, only {\em S65} shows different type of variability with $\D m = 0.0413$
amplitude and is, most probably, a true variable (Fig.~\ref{fig:S65}). It is interesting to note 
that star {\em S240}, apart from obviously
induced variability, displays the possible signature of a EA-type eclipsing 
binary light curve. However, to determine the true light curve, this star 
has to be examined in greater detail. This will be done in a separate study. 

\begin{deluxetable}{lccclll}
\tablecaption{Some of the standards from Stetson's catalogue (Stetson 2000) 
that exhibited spurious variability during our observational run. Column (1) 
is star's ID by Stetson (2000), (2) and (3) are mean standard $R$ magnitudes 
and observed variation, (4) is the determined period when possible. 
Columns (5), (6) and (7) are the computed variability statistics 
(Sec.~\ref{sec:criteria}): ${\cal A}$ is the alarm statistics, ${\cal F}$ 
is the significance level of periodicity and $\sigma_{XS}$ is the excess 
variance. \label{table:var-standards}} 
\tablewidth{0pt}
\tablehead{
\colhead{ID}  & \colhead{$\la R\ra$(mag)} & \colhead{$\Delta r$(mag)}   & 
\colhead{Period(d)}  &  \colhead{${\cal F}$} & \colhead{${\cal A}$} & \colhead{$\s_{\rm XS}$}
}
\startdata
S65 &  17.121  & 0.042  & 0.2172   &$1.0\times10^{-7} $&1.84& 0.04    \\
S70 &  15.567  & 0.182  & spurious &$6.2\times10^{-10}$&16.9& 0.22    \\
S72 &  17.151  & 0.085  & spurious &$6.4\times10^{-8} $&8.20 & 0.09    \\
S74 &  17.338  & 0.157  & spurious &$1.7\times10^{-8} $&3.83& 0.10     \\
S80 &  16.063  & 0.080  & spurious &$8.1\times10^{-11}$&13.8& 0.10     \\
S230&  15.427  & 0.050  & spurious &$2.4\times10^{-10}$&11.9& 0.07    \\
S240&  15.843  & 0.184  & spurious &$1.5\times10^{-8} $&14.4& 0.21    \\
\enddata
\end{deluxetable}

\section{Detection of New Variables}
\label{sec:new_variables}

The primary goal of our main survey is to identify possible microlensing 
events. We have used the commissioning data set to tune up and test our data 
reduction and analysis pipeline, without necessarily expecting to find new variable 
candidates in this data set. However, it is still useful to search for 
variability, since there is a chance of discovery and the search can 
also yield interesting variable stars not found through earlier variable 
selection methods. Furthermore, we would expect to find false positives in the 
microlensing search --- variables stars like dwarf novae, classic novae, etc (see 
\citet{SafonovaStalin} for discussion on contamination of ML searches). Finding 
these events and successfully identifying them would demonstrate our 
ability to detect actual microlensing events in our data set in the future.

To search the data set of $\sim 10,000$ light curves for variable stars, we take 
several main steps. Due to the large numbers, it is necessary to automate 
the detection of variable sources. We found that no single algorithm is 
appropriate for the detection of all kinds of variable stars present in our 
data, and that false positives (or missing variables) are high if we use just one 
algorithm. We thus implement several cuts to select promising candidates.
In addition, we only consider light curves having more than 60 data points, since  
several variability detection algorithm can produce wrong results
when significant amount of data is missing.

\subsection{Alarm Statistics}

As a first algorithm, we have selected the alarm statistics from
the VARTOOLS software package \citep{vartools}. The `alarm' $\cal A$ is defined 
as \citep{alarm_stat}
\be
{\cal A}=\fr{1}{\chi^2}\sum_{i=1}^{M}\left(\fr{r_{i,1}}{\s_{i,1}} + 
\fr{r_{i,2}}{\s_{i,2}} +
\ldots + \fr{r_{i,k_i}}{\s_{i,k_i}}\right)^2 - \left(1+\fr{4}{\pi}\right)\,,
\label{eq:alarm}
\ee
where $r_{i,j}$ is the residual of the $j$-th measurement of the $i$-th run and
$\s_{i,j}$ is its uncertainty. The sum is over all the measurements in a run
and then over the $M$ runs. A `run' is defined as a maximal series of consecutive 
residuals (in the folded light curve) with the same sign. The $\chi^2$ is the 
known function
\be
\chi^2=\sum_{i=1}^N\left(\fr{r_i}{\s_i}\right)^2\,,
\label{eq:chi^2}
\ee
where the sum is over $N$ observations. In contrast to $\chi^2$ itself, 
$\cal A$ is not sensitive to a systematic overestimation or underestimation 
of the uncertainties. It is easy to see that $\cal A$ is minimal 
when the residuals alternate between positive and negative values, 
and that long runs with large residuals increase its value. The minimal 
value of the summation is exactly $\chi^2$, and therefore the minimal 
value of $\cal A$ is $-4/\pi$. We find that as an initial assessment, 
the alarm statistics is good for large sudden (aperiodic) variations, 
but that it fails in case of the short-period variability. It cannot 
easily distinguish between non-variability with large noise and very 
regular short periodicity, which we already noticed when we tried to use 
the ordinary $\chi^2$-statistics. However, 
even in detecting large variations, the alarm results have to be viewed with 
caution, as it picks up the false variability, like in case of 
systematic variations found in standard stars (see Sec.~\ref{sec:stan-var}).
We still found it useful, as high values of alarm statistics may indicate 
eclipsing binary, microlensing event or any non-periodic transient.

\subsection{Excess variance method}

Any light curve of a variable star candidate varies due to measurement
errors $\sigma_i$ and intrinsic variations. The variance of such a light curve
consisting of $N$ data points with amplitude $X_i$ is given by
\begin{equation}
S^2 = \frac{1}{N-1} \sum_{i=1}^{N} \left(X_i - <X>\right)^2\,.
\end{equation}
This measured variance has contributions from both intrinsic source variability
and measurement uncertainty. Therefore to know if any intrinsic variations are
present in the light curve, one needs to remove the contribution of 
measurement errors to the observed variance. The commonly used approach to 
get an idea of the intrinsic variation present in the candidate light curve
is to use the `excess variance' \citep{{Nandra1997},{Vaughan2003}},
which is defined as 
\begin{equation}
\sigma^2_{XS} = S^2 - \sigma^{2}_{err}\,,
\end{equation}
where $\sigma^{2}_{err}$ is the average variance of the $N$ measurements, given as
\begin{equation}
\la \sigma^{2}_{err}\ra = \frac{1}{N} \sum_{i=1}^{N} \sigma^{2}_{err,i}\,.
\end{equation}

A large value of $\sigma^2_{XS}$, much in excess of the measurement errors
might hint for the presence of variations in the light curve.
 
\subsection{Lomb-Scargle Periodogram}
\label{sec:LS}

The Lomb-Scargle (LS) periodogram \citep{LS} is an algorithm designed to pick out 
periodic variables in an unevenly sampled data. The periodogram statistic, $\Theta$, 
measures the fit for a given pulsation frequency. The probability distribution of 
$\Theta$ is then used to calculate the probability, $P(\Theta > c)$ of obtaining 
the value of the periodogram higher than the actual observed value, $\Theta=c$, 
from a hypothetical pure noise signal. An unlikely good fit, corresponding to small
$P$, is interpreted as detection of the corresponding period. Its complement 
probability, ${\cal F}=1-P(\Theta>c)$, is called the significance level.
Thus, the likelihood of the existence of a 
periodic signal can be established with the `false alarm probability' ${\cal F}$ 
--- a simple estimate of the significance of the height of a peak in the periodogram. 
A small value of ${\cal F}$ indicates a highly significant periodic signal. 
We have used the implementation of this algorithm given in \citet{Press}
and this seems to be a good criteria to pick periodic variability present
in the dataset. 

\subsection{Final Selection}
\label{sec:criteria}

For selecting candidate variable stars from the original light curve
database, we used a combination of three algorithms described above, 
namely (a) alarm statistics, (b) excess variance and (c) Lomb
periodogram. After applying these three algorithms to the sets of 
secondary photometric standards and variable stars known in M53 
(total 114 stars), we have devised the following  criteria for 
the final selection of variables from our list of $\sim 9,700$ 
candidate variable stars, 
\berr
\begin{cases}
{\cal A} > 1.0 \,;\hspace{5cm}\\
{\cal F} < 10^{-4}\,;\\
\sigma_{XS} > 0.09\,;\\
\mbox{rms}>0.01\,.
\end{cases}
\label{eq:criteria}
\err
A total of  310  candidate variable stars were found to satisfy
the criteria simultaneously. For these candidates, we calculated
the periods using two algorithms: the Lomb periodogram and
the \citet{LK} algorithm (LK). From the highest peak in the
Lomb periodogram we detect the initial estimate of the period. This period 
is then passed on to LK algorithm. LK method provides a systematic approach 
to testing a series of trial 
periods (with an increment of 0.0001 days) and looking for the period that 
results in the "smoothest" phase 
curve. The $N$ observed points are sorted by phase, and the sum of the 
squares of the difference in magnitude of successive pairs of points is 
used to rank the trial period. The smallest value of the figure of merit 
\be
\Theta=\sum_i \left(m_i-m_{i+1}\right) \big/ \sum_i
\left(m_i-\overline{M}\right)^2\,,
\label{eq:lktheta}
\ee
where $\overline{M}=m_i/N$ should be the nearest to the correct period 
since this represents the smallest successive changes in the light curve.

\subsection{False Positives}
\label{sec:false}

Finally, we visually inspect the light curves to remove false 
positives. We noticed that due to the intra-night changes of
the PSF, several stars selected as variables by our selection
criteria, have, in fact, variability induced by the same mechanism as was 
discussed in Sec.~.\ref{sec:caveats}. This type of variability is  
easily detected by eye as it is virtually identical in such false light curves, 
and these stars were removed from the candidate list. We also noted that during
our continuous observations of $\sim 7.5$ hrs, the coordinates of the
field centre drift between the images through the night. The typical
intra-night drift was $\sim 25$ pixels, but it had some effects in our data.
The drift caused stars at the edges of the field to enter and exit the
CCD's field of view during the night, resulting in incomplete light curves for
those stars. We remedied that by eliminating the strips of $\sim 20$-pixel width
along the edges from our images. Finally, we reject any variable star that is 
less than 10 pixels away from a variable candidate and has higher rms in flux. 
This criterion enabled us to eliminate stars whose variability was induced by 
its proximity to a genuine variable.

\subsection{The Classification of New Variables}

The basic characteristics of our observations and the achieved precision of
the relative photometry allowed us to establish the fact of light 
variations in the variable star candidates. However, the total duration of our 
observations was not long enough for reliable classification of a large number 
of them. Nevertheless, the obtained light curves made it possible to 
tentatively estimate the type of some of the discovered variables from the
shape and characteristic features of the light curves.  

\subsubsection{Candidate SX Phe-type stars}
\label{subsec:SXPhe}

We have found 22 stars whose light curves parameters, i.e. short periods and
low amplitudes, allow us to classify them as potential SX Phe stars. SX Phe 
stars known in globular clusters usually have periods between $0.03 - 0.14$ days 
and they often show multiple frequencies of light variations. However, with our 
limited span of observations, it was very difficult to establish the complicated 
frequency patterns, so we aimed to find at least the main periodicity. The parameters 
of these candidates are listed in Table~\ref{table:SXPhe}. For each star we provide 
designation, equatorial coordinates ($\a,\delta$), period $P$, mean brightness in $R$, 
range of variability $\Delta r$, epoch of light-minimum $T=JD-2455280$ and in the 
Notes we present the BSS match, possible variability type, or pulsation mode as 
inferred from the $P-L$ diagram. 
 
\begin{deluxetable}{lcccclll}
\tablecaption{\small Equatorial coordinates and light curve parameters for 
new SXPhe-type candidates in M53. $\Delta r$ and epoch were 
obtained by IRAF task {\em pdm}.\label{table:SXPhe}} 
\tablewidth{0pt}
\tablehead{
\colhead{Variable} & \colhead{$\a$(2000)} & \colhead{$\delta$(2000)} &
\colhead{P}   & \colhead{$<R>$}      & \colhead{$\Delta r$} & 
\colhead{epoch}    & \colhead{Note} \\
\colhead{designation} & \colhead{h:m:s)} & \colhead{($^{\circ}:':''$)} &
\colhead{(d)}   & \colhead{(mag)}      & \colhead{(mag)} & 
\colhead{(d)}    & \colhead{ } 
}
\startdata
SX2      &13:12:48.29&18:13:18.7&0.039362 & 19.6018 & 0.2924  &0.2039&SXPhe,F,BSS  \\
SX3      &13:12:48.27&18:14:34.5&0.059824 & 18.7855 & 0.16501 &0.1763&SXPhe,F,BSS  \\
SX4      &13:12:48.69&18:10:10.2&0.049500 & 19.4494 & 0.41865 &0.3733&SXPhe,F,BSS  \\
SX5      &13:12:48.70&18:10:11.3&0.050400 & 19.3330 & 0.38424 &0.2634&SXPhe,F  \\
SX6      &13:12:49.91&18:08:56.5&0.044755 & 18.1171 & 0.16693 &0.3482&SXPhe,1H,BSS \\
SX7      &13:12:51.74&18:10:33.8&0.132701 & 19.6519 & 0.37991 &0.2409&RRl? \\
SX8      &13:12:52.03&18:09:53.6&0.099101 & 18.7416 & 0.37707 &0.2517&SXPhe,F      \\
SX9\_1$^{\dagger}$&13:12:52.91&18:10:35.7&0.054056&18.5738&0.2819   &0.216 &SXPhe,BSS?    \\
SX9$^{\dagger}$   &13:12:52.99&18:10:35.4&0.056756&19.6975&0.59367  &0.216 &BSS?    \\
SX11     &13:12:53.66&18:08:57.7&0.052932 & 19.3140 & 0.42881 &0.2807&SXPhe,F      \\
SX12     &13:12:53.89&18:09:13.2&0.100001 & 19.4074 & 0.32196 &0.3807&not SXPhe, uncertain  \\
SX13     &13:12:54.52&18:09:32.6&0.112712 & 19.4671 & 0.49331 &0.3733&not SXPhe, uncertain \\
SX14     &13:12:54.78&18:09:37.6&0.071130 & 18.6844 & 0.52035 &0.3032&SXPhe,F,BSS  \\
SX15     & 13:12:56.34 & 18:11:54.2  & 0.137101 & 20.4167 & 0.60844 &0.3258&not SXPhe, RRl?  \\
SX16     &13:12:56.37&18:11:05.4&0.049900 & 17.9848 & 0.17625 &0.1637&SXPhe,1H     \\
SX17     &13:12:57.17&18:09:41.9&0.040567 & 18.0077 & 0.13385 &0.1824&SXPhe,1H,BSS \\
SX24     &13:12:57.64&18:10:43.0&0.033724 & 18.4832 & 0.16544 &0.3733&SXPhe,1H     \\
SX19     &13:12:58.30&18:08:41.3&0.044378 & 19.3035 & 0.33948 &0.1637&SXPhe,F,BSS  \\
SX20     &13:12:59.53&18:09:17.5&0.134857 & 18.7493 & 0.16805 &0.384 &  RRl?       \\
SX21     &13:12:59.51&18:11:17.4&0.037762 & 19.5682 & 0.26318 &0.2983&SXPhe,F,BSS  \\
SX22     &13:13:01.92&18:12:30.4&0.045955 & 19.5604 & 0.38023 &0.1824&SXphe,F      \\
SX23     &13:13:04.21&18:10:59.4&0.107312 & 19.6818 & 0.38479 &0.2039&not SXPhe, uncertain\\
SX25     &13:12:57.37&18:10:15.3&0.066868 & 19.1302 & 0.96035 &0.4111&$\equiv$V80\_5,SXPhe,F      \\
\enddata
\tablecomments{$^{\dagger}$ These two stars are separated by $1^{\as}.2$ 
arcsec. Though their periods and epochs are nearly the same which would 
argue that there only one is variable, the amplitude of a fainter star 
is twice that of a brighter star. They also match within $1^{\as}$ to 
a BSS (see Sec.~\ref{sec:BSS}), which coordinates \citep{Beccari08} 
are located exactly between these two stars. Their BSS nature is 
also under question, as \citet{Beccari08} could not resolve them 
and thus must have used their combined light to derive their 
conclusion. More discussion is in the text.}
\end{deluxetable}

Several methods are available to confirm the genuine nature of the SX Phe 
candidates. It is known that observational identification of the pulsational 
modes in SX Phe is difficult. \citet{McNamara} and, subsequently  
\citet{Rodriguez}, suggested that at fundamental mode, SX Phe 
preferably show large amplitudes ($>0.2$) and asymmetric light curves. 
They also follow a tight relation between their 
fundamental mode period and luminosity ($P-L$ relation). SX Phe are also the 
blue straggler stars found in a large number in globular clusters ($\sim 200$ 
in this cluster, Beccari et al. 2008). Thus, a match between known BSS and 
our SX Phe  candidates would argue in favour of their SX Phe nature (see Sec.~\ref{sec:BSS}).

\begin{figure}[ht!]
\begin{center}
\includegraphics[width=7cm,height=7cm]{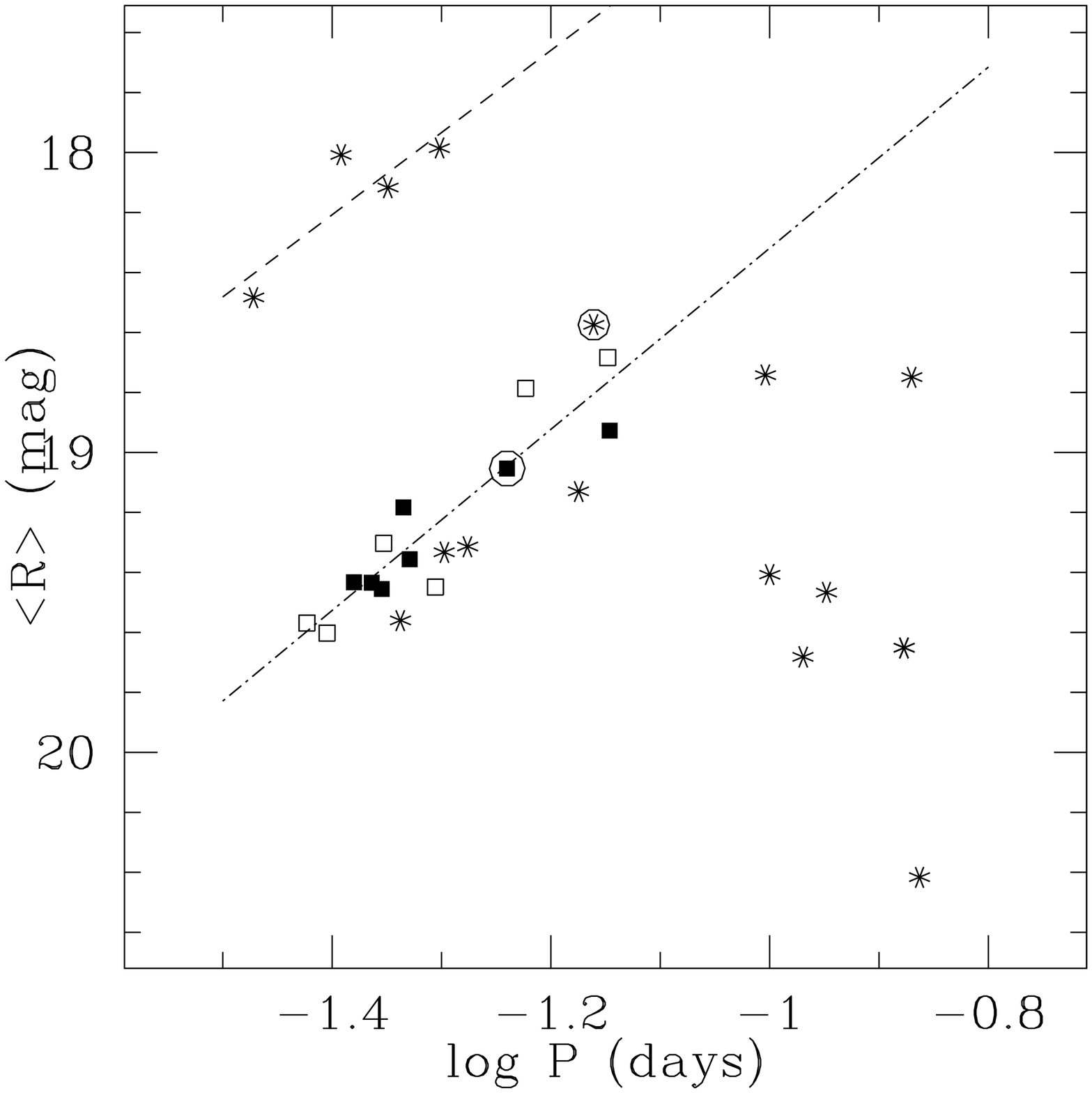}
\caption{Mean $R$ magnitude vs. period diagram. Squares represent stars used 
in the fundamental mode fitting; filled squares are previously known SX Phe in M53
and open squares are new SX Phe. Filled circled square is the star {\em V74} 
with the period converted from first harmonic to fundamental mode. Asterisks 
are remaining SX Phe candidates. Circled asterisk is an SX Phe candidate 
{\em SX9\_1} with the period converted to fundamental mode. Four top asterisks 
are most probably SX Phe in a higher pulsational mode and we have derived the linear 
fit for them.\label{fig:PL}}
\end{center}
\end{figure}

Based on a study of 6 SX Phe stars in M53, \citet{Jeon03} derived the following
$P-L$ relation for the fundamental mode
\be
<V>=-3.01(\pm 0.262) \log{P} + 15.31(\pm 0.048)\,.
\label{eq:jeon}
\ee
We have used our derived periods and $R$ magnitudes for 7 previously known 
SX Phe stars ({\em V73, V74, V75, V76, V87, V89} from \citet{Jeon03} and 
{\em V79} from DK) and 6 new SX Phe candidates that have BSS matches 
(except {\em SX9, SX9\_1}, {\em SX6} and {\em SX17}, see the discussion 
later in this subsection) to derive the $P-L$ relation. Our relation is 
surprisingly nearly identical to that of \citet{Jeon03},
\be
<R>=-3.019(\pm 0.382) \log{P} + 15.300(\pm 0.501)\,.
\label{eq:ourfit}
\ee
In Figure~\ref{fig:PL}, we display the period and mean $R$ magnitude relation 
for 7 known SX Phe stars and 22 new SX Phe candidates in M53. We note that variable 
{\em V74} was found by \citet{Jeon03} to lie far away from 
the $P-L$ line; this is possibly due to the wrong period reported by the authors 
in their Fig.~7. Besides, the authors excluded this variable from the 
fundamental $P-L$ fitting suggesting that it was in a higher pulsational mode. However, if it is in the first harmonic mode, its period can be converted to 
the fundamental mode by the theoretical relation $P_{1H}/P_F=0.783$ 
\citep{Jeon04}. Once corrected for that, 
we can see that {\em V74} lies on the $P-L$ line together with other SX Phe stars 
(see Fig.~\ref{fig:PL}). In our sample of SX Phe candidates we have two 
stars which have similar periods but different magnitudes (variables {\em SX9\_1} 
and {\em SX\_9} in Table~\ref{table:SXPhe}). It is possible that only 
one of them is variable and another has variability induced by its proximity. 
By the selection criteria adopted in Sec.~\ref{sec:false}, we choose the star 
which has smaller rms in flux to be genuinely variable; thus, we select the star 
{\em SX9\_1}. This star lies above the fundamental mode $P-L$ line; when we convert 
its period by using the same relation $P_{1H}/P_F=0.783$ \citep{Jeon04}, it falls 
close to that line. Stars with converted periods, {\em SX9\_1} and {\em V74}, 
are marked by circles in the Figure~\ref{fig:PL}.
  
\begin{figure*}[ht!]
\begin{center}
\includegraphics[width=17cm,height=17cm]{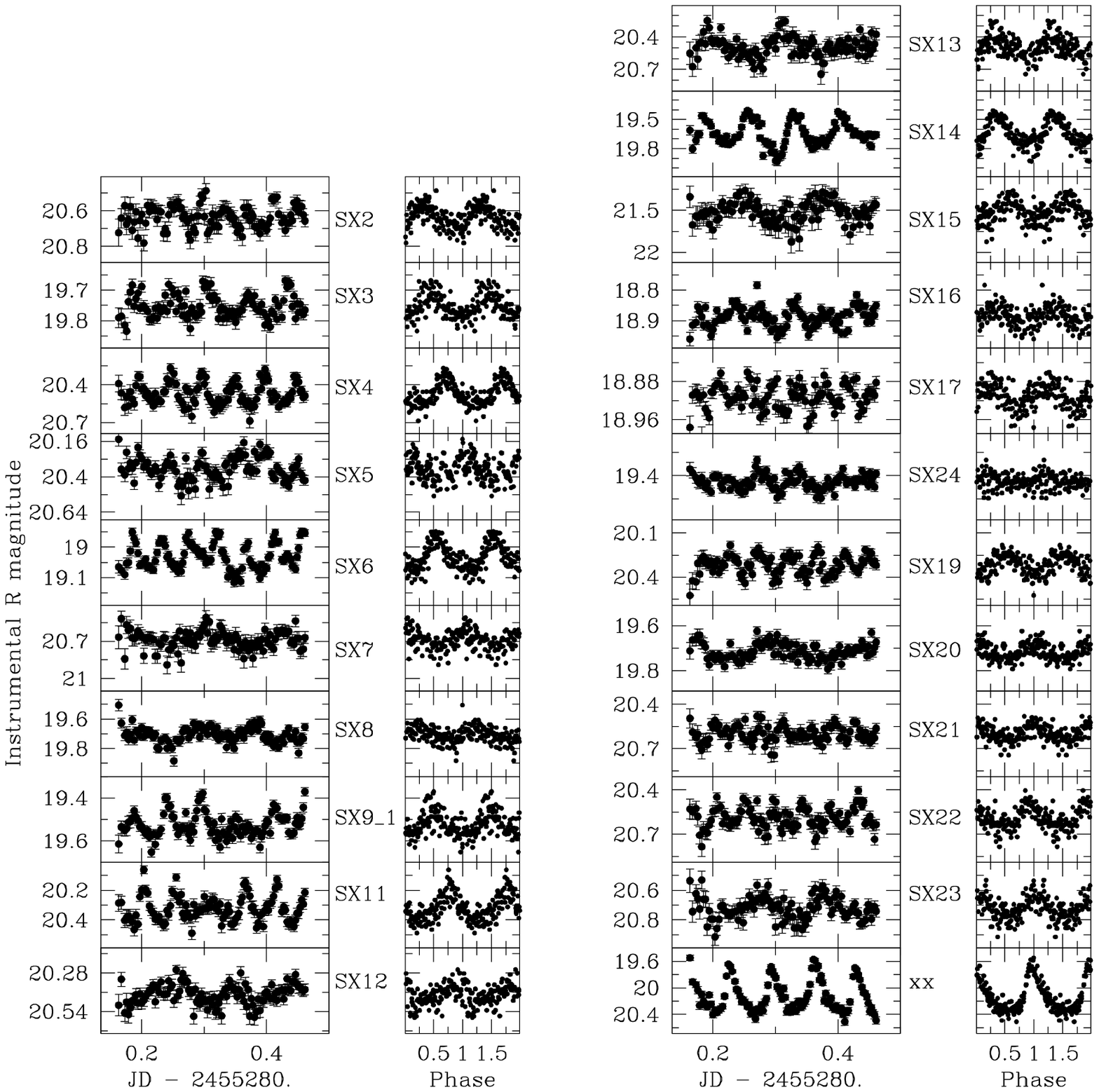}
\vskip -0.3in
\caption{Phased and time-domain light curves of 22 SX Phe candidates from 
Table~\ref{table:SXPhe}.\label{fig:SX-lc}}
\end{center}
\end{figure*}

The stars that fall along the $P-L$ relation can be considered with high probability, 
given also their periods and shape of light curves, as belonging to the SX Phe type. 
Stars in the top left quadrant of the plot (variables {\em SX6, SX16, SX17, SX24}) 
are presumably in a higher pulsational mode. They also have smaller amplitudes 
(0.1606 on average) comparing to the ones pulsating presumably in fundamental mode 
(0.4365 on average). We have derived the least square fit for them and found that it 
is nearly parallel to the fundamental mode fit,
\be
<R>=-2.740(\pm 1.160) \log{P} + 14.371(\pm 1.600)\,.
\label{eq:secondfit}
\ee
Stars that are located to the right and below the fundamental $P-L$ line though 
undoubtedly variable (see their light curves in Fig.~\ref{fig:SX-lc}), are most 
probably not SX Phe type. Stars {\em SX7} and {\em SX20} are possibly RR Lyrae, 
while others ({\em SX12, SX13, SX15, SX23}) are of uncertain type. But since their 
periods exceed 0.1 days, we cannot truly determine their nature. We thus confirm SX 
Phe-type fundamental mode pulsations for 11 new variables (namely, 
{\em SX2, SX3, SX4, SX5, SX8, SX10, SX11, SX14, SX21, SX22, SX25}) and 
probable higher pulsational mode for 5 new variables: {\em SX6, SX9\_1, SX16, 
SX17, SX24}. The phase and time-domain light curves of all 22 stars are shown 
in Figure~\ref{fig:SX-lc} and their parameters are given in 
Table~\ref{table:SXPhe}. The positions of 16 confirmed SX Phe stars on 
the CCD image are given in Figure~\ref{fig:SX}.

Given the results of our study, the number of confirmed SX Phe stars in M53 
reaches 28: 12 previously reported (with exclusion of {\em V81}, 
{\em V82} and {\em V83}) plus 16 discovered by us, and it is quite 
possible that there are much more yet undiscovered. SX Phe stars in globular
clusters can have very short periods (down to 0.025 days) and very small 
amplitudes, for ex., 25\% of 149 SX Phe stars in \citet{Rodriguez} catalogue 
have amplitudes of $< 0.^{m}05$. Thus we cannot exclude the possibility that many
of the apparently non-varying stars in the SX Phe region vary, but with 
undetectable by us amplitudes. More sensitive observations are necessary to 
search for this exciting class of variables. 

\subsubsection{RR Lyrae Candidates}
\label{subsec:RRl}

\begin{figure}[h!]
\begin{center}
\hspace*{-1.0cm} 
\includegraphics[width=9cm,height=12cm]{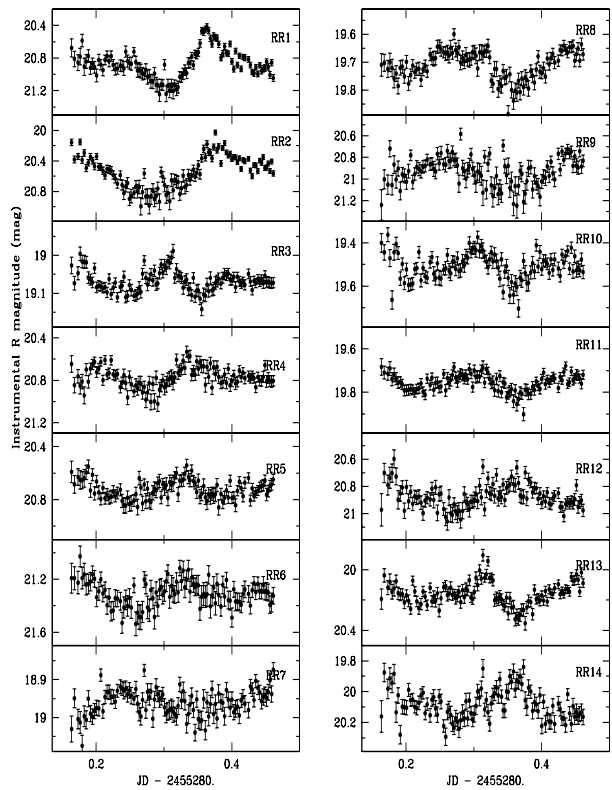}
\vskip -0.3in
\caption{Time-domain light curves of RR Lyrae candidates.\label{fig:RR-lc}}
\end{center}
\end{figure}

We tentatively classify the variable candidates with estimated periods of 
$ 0.15 < P < 0.2$ days and a characteristic shape of a light curve as RR 
Lyrae variables \citep{Clement}. The light curves of these 14 RR Lyrae 
candidates are shown in Figure~\ref{fig:RR-lc}. The equatorial coordinates 
($\a,\delta$) of these stars, mean brightness in $R$, rough estimate of the period, 
amplitude of variability $\Delta r$, the epoch of light-minimum ($T=JD-2455280$) 
and a possible variability type are given in Table~\ref{table:RRl}. Their positions 
on the CCD image are given in Figure~\ref{fig:uncertainRF} {\it Top}.

\begin{deluxetable}{lcccllll}
\tablecaption{\small Equatorial coordinates and light curve parameters for new RR 
Lyrae-type stars in M53. $\Delta r$ and epoch were obtained 
by the IRAF task {\em pdm}.\label{table:RRl}}
\tablewidth{0pt}
\tablehead{
\colhead{Variable}   & \colhead{$\a$(2000)}  & \colhead{$\delta$(2000)} &
\colhead{$<R>$}      & \colhead{Period}             & 
\colhead{ $\Delta r$}& \colhead{epoch}      &  \colhead{Remarks} \\
\colhead{designation}   &\colhead{(h:m:s)} & \colhead{($^{\circ}:':''$)} &
\colhead{(mag)}      & \colhead{(d)}             & 
\colhead{ (mag)}& \colhead{(d)}      &  \colhead{ } 
}

\startdata
RR1        &13:12:45.40&18:09:04.1  & 19.809   & 0.071   &0.82294 & 0.2933&  RRab      \\
RR2        &13:12:48.74&18:10:12.7  & 19.524   & 0.5034  &1.04199 & 0.2782&  RRab       \\
RR3        &13:12:53.82&18:09:18.7  & 18.151   & 0.2924  &0.15097 & 0.3556&  RRc       \\
RR4        &13:12:54.03&18:09:11.4  & 19.720   & 0.071   &0.48068 & 0.2909&  RRc        \\
RR5        &13:13:01.33&18:10:15.9  & 19.682   & 0.5034  &0.31814 & 0.3807&  RRc       \\
RR6        &13:12:47.13&18:10:29.0  & 20.215   & 0.2924  &0.48075 & 0.2382&  RRc?      \\
RR7        &13:12:54.58&18:09:45.4  & 18.074   & 0.071   &0.21463 & 0.1793&  RRc?          \\
RR8        &13:12:45.18&18:11:22.4  & 18.739   & 0.5034  &0.32282 & 0.3507&  RRc       \\
RR9        &13:12:48.29&18:13:18.7  & 19.846   & 0.2924  &0.84712 & 0.1637& RRc?        \\
RR10       &13:12:38.97&18:09:07.2  & 18.556   & 0.071   &0.3408  & 0.3659& RRab?         \\
RR11       &13:12:45.18&18:11:22.4  & 18.788   & 0.5034  &0.2573  & 0.3733&  RRc       \\
RR12       &13:12:48.29&18:13:18.7  & 19.820   & 0.2924  &0.46049 & 0.261 &  RRc       \\
RR13       &13:12:38.97&18:09:07.2  & 19.156   & 0.071   &0.45296 & 0.3757&  RRab      \\
RR14       &13:12:45.18&18:11:22.4  & 19.090   & 0.5034  &0.45335 & 0.2585&  RRc    \\
\enddata
\tablecomments{The periods are derived from incomplete
light curves and are only approximate due to the short span of our observations.} 
\end{deluxetable}

\subsubsection{Eclipsing Binaries Candidates}
\label{subsec:WUMa}

\begin{figure}[hb!]
\begin{center}
\hspace*{-1.0cm}\includegraphics[width=12cm,height=15cm]{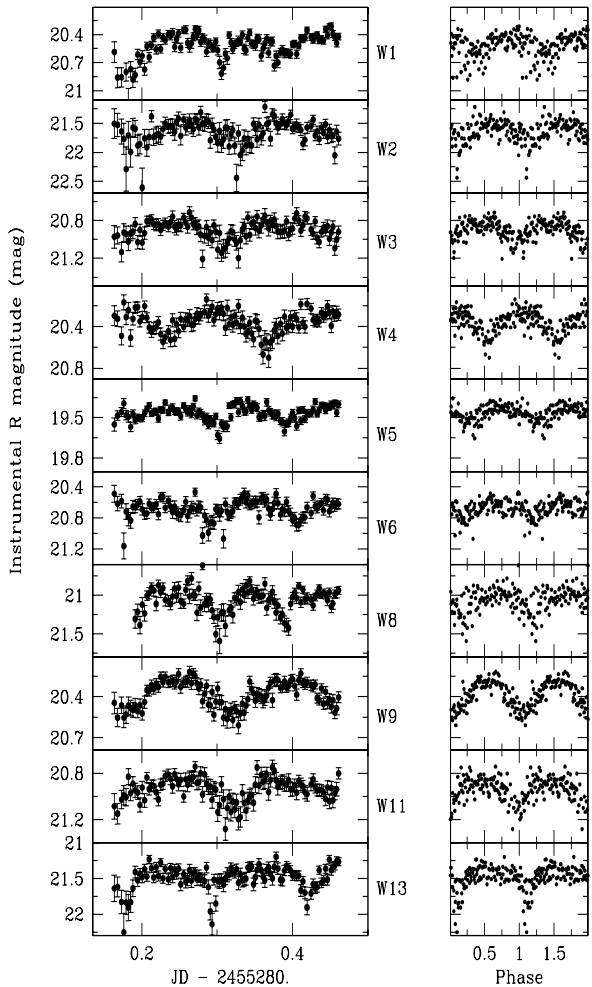}
\vskip -0.3in
\caption{Phase and time-domain light curves of W UMa candidates.\label{fig:W-lc}}
\end{center}
\end{figure}

\begin{figure}
\begin{center}
\vspace*{-2.0cm}
\includegraphics[width=8cm,height=7cm]{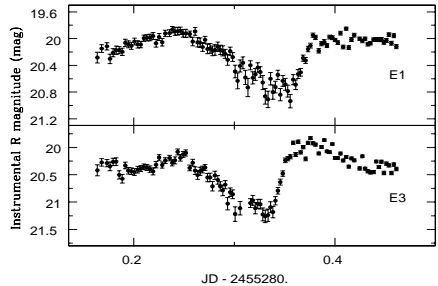}
\vskip -0.2in
\caption{Light curves of long-period eclipses.\label{fig:eclipse}}
\end{center}
\end{figure}

Though about $14\%$ of the nearly 200 BSS in this cluster are estimated to 
be in binary systems \citep{Beccari08}, no eclipsing binary was
ever found. Here we report the discovery of 10 stars of W Ursa
Majoris (W UMa) type and 2 detached eclipsing binaries. W UMa-type stars are 
contact binaries and it was noted (Kaluzny 1997) that almost all known 
eclipsing main-sequence binaries with periods shorter than about 0.4 days show
EW-type light curves. All our suspected W UMa-type candidates have periods 
of $0.1$ d on average and indeed mostly display EW-type light curves. For each 
star we provide designation, equatorial coordinates ($\a,\delta$), mean brightness 
in $R$, rough estimate of the period, range of variability $\Delta r$, epoch 
of light-minimum $T=JD-2455280$ and in the Notes we present a possible 
variability type or BSS match. The phase and time-domain curves are presented 
in Figure~\ref{fig:W-lc}. In Figure~\ref{fig:eclipse} we show the time-domain 
light curves of two long-period eclipses {\em E1} and {\it E3}, which can be either 
detached binaries or eclipsing Cataclysmic Variables (CV). The positions of these 
binaries on the CCD image are given in Figure~\ref{fig:uncertainRF}.

\begin{deluxetable}{lcccclll}
\tablecaption{\small Equatorial coordinates and light curve parameters for new 
candidate eclipsing binaries in M53. $\Delta r$ and epoch were obtained 
by the IRAF task {\em pdm}.\label{tab:WUMa}} 
\tablewidth{0pt}
\tablehead{
\colhead{Variable}  & \colhead{$\a$(2000)} &  \colhead{$\delta$(2000)} &
\colhead{Period}  & \colhead{$<R>$ }  &
\colhead{$\Delta r$} & \colhead{epoch} & \colhead{Remarks}  \\
\colhead{designation}   &\colhead{(h:m:s)} & \colhead{($^{\circ}:':''$)} &
\colhead{(d)}      & \colhead{(mag)}             & 
\colhead{ (mag)}& \colhead{(d)}      &  \colhead{ } 
}
\startdata
W1      &13:12:43.70&18:10:09.0&0.063701& 19.495  &  0.79939&  0.1824& EW?   \\
W2      &13:12:51.07&18:11:11.9&0.152667& 20.522  &  1.22333&  0.3258& EW?    \\
W3      &13:12:51.41&18:09:37.4&0.039362& 19.846  &  0.4898 &  0.2807& EW?   \\
W4      &13:12:53.98&18:09:49.0&0.063701& 19.339  &  0.7531 &  0.3683& EW?   \\
W5      &13:12:53.21&18:09:47.3&0.152667& 18.520  &  0.29791&  0.3032& EW?    \\
W6      &13:12:57.46&18:10:29.6&0.039362& 19.641  &  0.60217&  0.3085& EW?    \\
W8      &13:12:56.66&18:08:18.7&0.063701& 19.999  &  0.97774&  0.3032& EW?   \\
W9      &13:12:59.00&18:10:21.9&0.152667& 19.369  &  0.39071&  0.3282& BSS, AH Vir?\\
W11     &13:13:01.04&18:10:01.6&0.039362& 19.881  &  0.53721&  0.311 & EW?\\
W13     &13:13:11.43&18:10:34.2&0.063701& 20.383  &  1.05576&  0.1763& EW?  \\
E1      &13:12:55.07&18:09:27.9&0.21427 & 19.203  &  1.08425&  0.3556& CV/EB? \\
E3      &13:12:59.98&18:09:24.8&0.24482 & 19.439  &  1.81670&  0.3085&BSS, EW/EB?\\
\enddata
\tablecomments{The periods are derived from incomplete
light curves and are only approximate due to the short span of our observations.}
\end{deluxetable}

\subsubsection{Unclassified Candidates}
\label{subsec:uncertain}

\begin{figure*}
\begin{center}
\hspace*{-1.5cm}
\makebox{
\includegraphics[width=16cm,height=17cm]{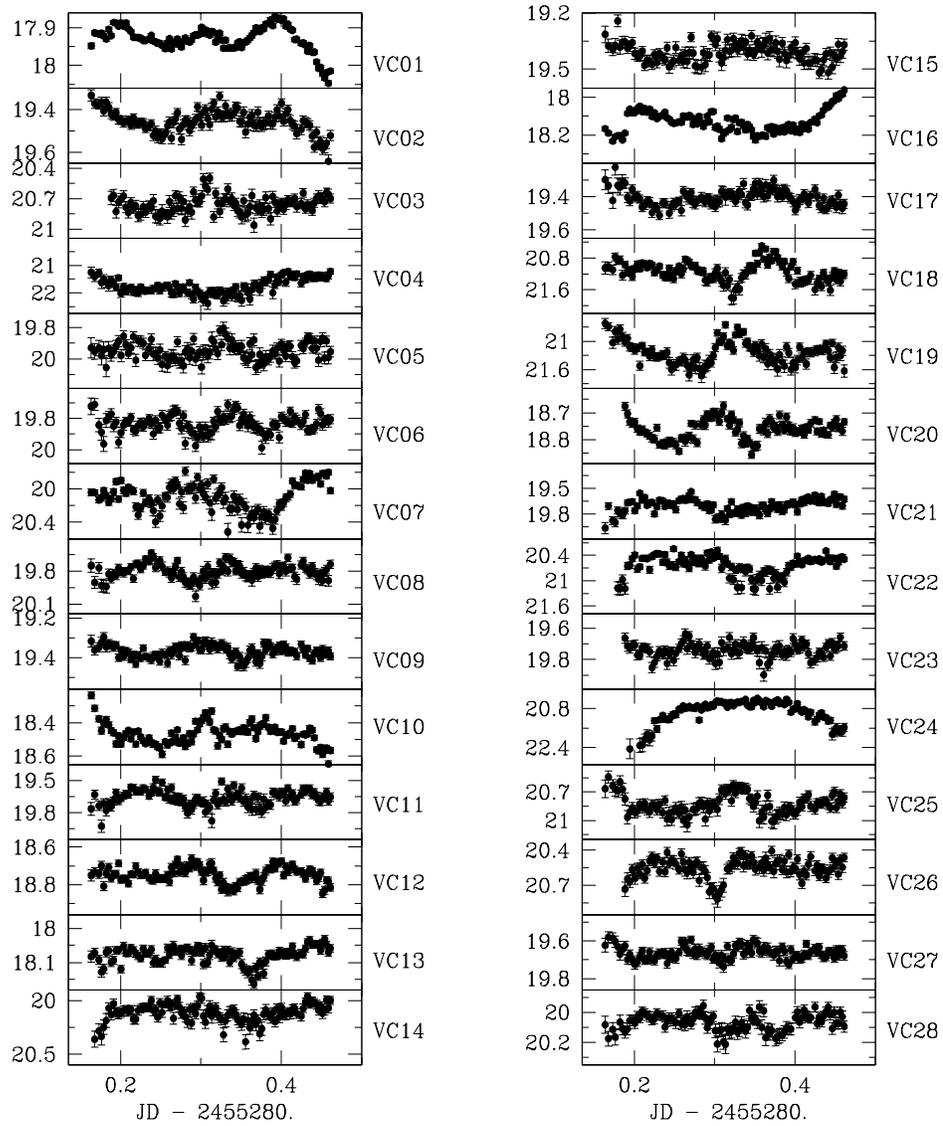}}
\vskip -0.3in
\caption{Light curves of uncertain type variables labelled with their designations
as given in Table~\ref{table:uncertain}.\label{fig:uncertain}}
\end{center}
\end{figure*}

In addition to the unclassified variable candidates discussed previously in
Sections \ref{subsec:V57}, \ref{subsec:V80} and \ref{subsec:SXPhe}, 
we have found 28 more stars that show definite light variations, but which are also 
impossible to classify due to the short span of our observations and their 
possible long periods and irregular nature. Their variability, however, is undoubtful judging from their light curves (Figure~\ref{fig:uncertain}). 
The equatorial coordinates and photometric parameters of these variables 
are presented in Table~\ref{table:uncertain}. Their positions on the CCD 
image are given in Figure~\ref{fig:uncertainRF}. Several of these stars 
were matched to the BSS catalogue by \citet{Beccari08} (Table~\ref{tab:BSS}). 
Star {\em VC23} is a possible long period variable (LPV), following the 
method described in \citet{Hartman} in which a star is defined as an 
LPV when a light curve can be fitted by a parabola. Also in Table~\ref{table:uncertain} is given the data on {\em V57\_3} and {\em V80\_7}.

\begin{deluxetable}{lcccccl}
\tablecaption{\small Equatorial coordinates ($\a,\delta$) and light curve parameters 
for new unclassified variable stars in M53.\label{table:uncertain}} 
\tablewidth{0pt}
\tablehead{
\colhead{Variable}   & \colhead{ $\a$(2000)}  & \colhead{$\delta$(2000)} & 
\colhead{Period}  & \colhead{$<R>$}  & \colhead{Remarks} \\
\colhead{designation}   &\colhead{(h:m:s)} & \colhead{($^{\circ}:':''$)} &
\colhead{(d)}           & \colhead{(mag)}&  \colhead{ }  
}
\startdata
VC1   &13:12:43.85 &18:10:13.0 & 0.100201 & 17.099 &    RRl?                 \\
VC2   &13:12:50.58 &18:10:18.9 & 0.207101 & 18.511 &                    \\
VC3   &13:12:50.66 &18:09:39.3 & 0.121557 & 19.713 &   BSS acs\#201010     \\
VC4   &13:12:50.96 &18:10:27.3 & 0.298433 & 20.637 &                     \\
VC5   &13:12:52.26 &18:10:21.3 & 0.118212 & 18.967 &                    \\
VC6   &13:12:52.45 &18:09:37.3 & 0.089110 & 18.861 &                    \\
VC7   &13:12:52.50 &18:10:13.6 & 0.147758 & 19.105 &       RRl?              \\
VC8   &13:12:52.83 &18:10:00.8 & 0.097510 & 18.835 &                    \\
VC9   &13:12:53.36 &18:09:31.6 & 0.110812 & 18.430 &                    \\
VC10  &13:12:53.61 &18:09:54.4 & 0.210301 & 17.597 &                     \\
VC11  &13:12:54.44 &18:10:31.2 & 0.096401 & 18.703 &                    \\
VC12  &13:12:54.76 &18:10:10.4 & 0.124401 & 17.855 &                    \\
VC13  &13:12:55.32 &18:10:21.7 & 0.177668 & 17.237 &    BSS pc\#100221       \\
VC14  &13:12:55.42 &18:10:39.0 & 0.190400 & 19.127 &                    \\
VC15  &13:12:55.71 &18:09:57.0 & 0.188068 & 18.470 &    BSS pc\#103181      \\
VC16  &13:12:55.84 &18:10:35.5 & 0.233201 & 17.284 &                     \\
VC17  &13:12:56.05 &18:09:59.5 & 0.185700 & 18.463 &    BSS pc\#103149      \\
VC18  &13:12:56.38 &18:10:52.1 & 0.095901 & 20.072 &                    \\
VC19  &13:12:56.43 &18:10:57.7 & 0.149258 &20.155  &\\
VC20  &13:12:56.59 &18:10:47.5 & 0.089911 &17.869  &\\
VC21  &13:12:56.93 &18:10:30.1 & 0.180000 &18.741  &\\
VC22  &13:12:56.95 &18:11:02.1 & 0.202601 &19.666  &\\
VC23  &13:12:57.09 &18:10:15.8 & 0.065368 &18.776  & LPV\\
VC24  &13:12:57.66 &18:10:50.0 & 0.298433 &19.978  &\\
VC25  &13:12:58.02 &18:09:06.8 & 0.157258 &19.777  &\\
VC26  &13:12:58.03 &18:09:03.1 & 0.115912 &19.530  &W UMa?\\
VC27  &13:12:58.07 &18:09:23.5 & 0.083209 &18.702  &\\
VC28  &13:12:53.09 &18:10:08.7 & 0.065629 &19.080  &SXPhe? BSS acs\#1200729 \\
V57\_3 &13:12:57.56 &18:10:12.8 & 0.1800   & 16.7022&BSS pc\#102387\\
V80\_7 &13:12:55.53 &18:09:56.9 & 0.104911 &19.575  & \\
\enddata
\tablecomments{The periods are derived from incomplete
light curves and are only approximate due to the short span of our 
observations.}
\end{deluxetable}

\subsection{Blue Stagglers in M53}
\label{sec:BSS}

We have cross-correlated the coordinates of the known and newly discovered 
variables with the BSS database of M53 by Beccari et al. 2008 (via private 
communication) and found 23 matches within 0.1 arcsec. Seven cases are matches 
with previously known variables, of which four, namely 
{\em V57}, {\em V75}, {\em V87} and {\em V89}, are reported by us for the first time 
(Table~\ref{table:var-coos}), and 16 are new. Out of them, ten are SX Phe
stars, one is a W UMa candidate, one is a long-period
eclipse and four are of uncertain type. All stars matched by us with BSS are 
shown in Table~\ref{tab:BSS}. For each 
star we provide BSS catalogue number by \citet{Beccari08} nomenclature,
our designation, mean $V$ magnitude from BSS catalogue, mean brightness in $R$
and period determined in this work, possible variability type and in the 
Remarks we indicate who has first reported the match. The study in which 
these new BS stars are analyzed by combining both \citet{Beccari08} and 
our catalogues is in progress.

\begin{deluxetable}{llccccl}
\tablecaption{\small Blue stragglers in M53. Column (1) is the star number from
BSS catalogue \cite{Beccari08}. For previously known SX Phe we 
retained the usual nomenclature \citep{Clement}. New BSS are are designated as 
follows: SX means SX Phe-type stars, W means W UMa-type stars, E stands for eclipse 
and VC stands for Variable Candidate (type yet unidentified).\label{tab:BSS}} 
\tablewidth{0pt}
\tablehead{
\colhead{Cat\#BSS}   & \colhead{Variable} & \colhead{$V_{\rm Bec}$} & 
\colhead{$<R>$}      & \colhead{$P_{\rm LK}$} & \colhead{Type}  & 
\colhead{Remarks} \\
\colhead{ }   &\colhead{ } & \colhead{(mag)} &
\colhead{(mag)}      & \colhead{(d)}             & 
\colhead{ }& \colhead{ } 
}
\startdata
lbc\#208210  & V74  & 19.2848 &   19.054   & 0.045055  & SXPhe   &first report by DK\\
lbc\#219783  & V75  & 19.5750 &   19.455   & 0.044178  & SXPhe   & new match  \\
lbc\#213681  & V76  & 19.8584 &   19.434   & 0.041467  & SXPhe   &first report by DK \\
lbc\#231344  & V79  & 19.7637 &   19.183   & 0.046255  & SXPhe   &first report by DK\\
lbc\#208804  & V89  & 19.6853 &   19.435   & 0.043278  & SXPhe   &  new match  \\
acs\#102387  & V87  & 18.8752 &   19.356   & 0.046855  & SXPhe   &  new match   \\
pc\#102387   & V57\_3& 18.8752 &  16.7022  &  0.1784   & ?       & see Sec.~\ref{subsec:V57}\\
acs\#200309  & SX4  & 19.2108 &   19.4494  & 0.049500  & SXPhe   &  new    \\
acs\#200279  & SX6  & 19.1799 &   18.1171  & 0.044755  & SXPhe   &  new   \\
acs\#200174  &SX9\_1& 19.0312 &   18.5738  & 0.054056  & SXPhe   &  new    \\
acs\#1200685 & SX14 & 18.3878 &   18.6844  & 0.071130  & SXPhe   &  new         \\
acs\#100512  & SX17 & 19.3060 &   18.0077  & 0.040567  & SXPhe   &  new      \\
acs\#100732  & SX19 & 19.5009 &   19.3035  & 0.044378  & SXPhe   &  new      \\
acs\#100681  & SX21 & 19.4517 &   19.5682  & 0.037762  & SXPhe   &  new     \\
lbc\#238275  & SX2  & 19.8489 &   19.6018  & 0.039362  & SXPhe   &  new       \\
lbc\#240911  & SX3  & 18.9840 &   18.7855  & 0.059824  & SXPhe   &  new      \\
acs\#100496  & W9   & 19.2661 &   19.369   & 0.152667  &WUMa?    &  new      \\
acs\#101000  & E3   & 19.6599 &   19.439   &  0.24482  &EW/EB?   &  new   \\
acs\#201010  & VC3  & 19.7295 &   19.713   & 0.121557  &?        &  new \\
acs\#1200729 & VC28 & 18.5123 &   19.080   & 0.065629  & ?       &  new  \\
pc\#100221   & VC13 & 19.5517 &  17.237    & 0.177668  &?        &  new   \\
pc\#103181   & VC15 & 19.1203 &  18.470    & 0.188068  & ?       &  new  \\
pc\#103149   & VC17 & 19.0852 &  18.463    & 0.185700  & ?       &  new   \\
\enddata
\tablecomments{$P_{\rm LK}$ means the 
period was obtained by LK method (see Sec.~\ref{sec:criteria}) and in some 
cases is only approximate due to the short span of our observations.}
\end{deluxetable}

\section{Conclusion}
\label{sec:conclusion}

We carried the all-night observation of the globular cluster NGC 5024 (M53)
using 2-m {\it HCT} telescope of the Indian institute of Astrophysics, Hanle,
India. As a result, we obtained 101 $R-$band images of the cluster. The reduction
of our data using the pre-released version of the DanDIA pipeline \citep{bramich08}
revealed significant short-term brightness variations in $\sim 300$ stars. 
79 of them are candidates for new variables, out of which 16 are SX Phe stars 
(our conclusion was based on the periods, shapes of light curves and analysis 
of $P-L$ relation), 10 are W UMa-type eclipsing binaries (the conclusion 
is based on the shapes of the light curves), 14 are RR Lyrae candidates 
(the conclusion is based on the shapes of the light curves) and 2 are 
long-period eclipses. The remaining 37 are of yet uncertain type due to 
our short observational run; because of the limited span of observations, 
the reliable determination of periods was limited to periods of $\leq 0.1$ days.  
Our results constitute the largest population of variables 
discovered in one research work on M53. Compared to the latest photometric 
study of this cluster by \citet{DK09}, our photometry depth reaches 21 magnitude 
in $R$, has better precision and better resolution. Consequently, 
we were able to identify many more new variables and, in a few cases, 
clarify the status of previously reported ones. We have shown that some 
previously known variables are mis-identified, and some are even non-variable. 
We also have shown that stars previously thought to be photometrically 
stable \citep{Stetson2000} can display short-period variability and that status 
of some of them as photometric standards possibly have to be revised. We refined 
the equatorial \coos for all previously known variables in our FOV and provided the 
equatorial coordinates and finding charts for newly discovered variable candidates.
We also report the error in the listing of the M53 variables coordinates in 
the online catalogue by Samus.

Our study shows that M53 has a number of variable stars comparable to M15, 
the most populous cluster in variables; however, we feel that we have 
barely scratched the surface. Extensive tests would be necessary to answer
the question about the number of variables which were missed. We are confident 
that with the advent of new differential imaging technique, using 2-m 
class telescopes and regular observations with moderate cadence, it will be 
possible to discover many more variables in this cluster.

\bibliographystyle{apj}
\bibliography{variability}

\end{document}